\documentclass[article]{nojss}

\usepackage{orcidlink,thumbpdf,lmodern}
\newcommand{\hide}[1]{}
\usepackage{amsmath,mathtools,latexsym,mathtools,bm,amsfonts}
\usepackage[mathscr]{eucal}
\usepackage{amssymb}
\usepackage{array}
\usepackage{enumitem}
\usepackage{longtable}
\usepackage{subcaption}
\usepackage{fancyvrb}
\usepackage{booktabs}
\usepackage{framed}
\usepackage{pifont}

\usepackage{tikz}
\usetikzlibrary{arrows.meta, positioning, shapes}
    
\usetikzlibrary{bayesnet}
\usetikzlibrary{fit,positioning}

\usepackage{pifont}

\author{Cornelius Fritz~\orcidlink{000000027781223X}\\School of Computer Science and Statistics\\Trinity College Dublin\\Ireland
   \And Michael Schweinberger~\orcidlink{0000000336495386}\\Department of Statistics\\The Pennsylvania State University\\USA}
\Plainauthor{Cornelius Fritz, Michael Schweinberger}

\title{\proglang{R} Package \proglang{iglm}:\\
Regression under Interference\\
in Connected Populations}
\Plaintitle{Regression under Interference}
\Shorttitle{Regression under Interference}

\Abstract{
We introduce \proglang{R} package \pkg{iglm},
which implements a unifying regression framework for studying relationships among binary, 
count-,
and real-valued predictors and outcomes in connected populations with interference due to spillover,
contagion,
and other complex phenomena.
On the computational side,
the regression framework provides scalable methods that can be applied to small and large data sets,
by solving a convex optimization program based on pseudo-likelihoods via accelerated Minorization-Maximization algorithms.
On the statistical side,
the regression framework comes with provable theoretical guarantees for statistical estimators,
including consistency results and rates of convergence for maximum pseudo-likelihood estimators.
To increase the versatility of \pkg{iglm},
users can add custom-built model terms.
We showcase \pkg{iglm} using data on hate speech on the social media platform X and communications among students.
}
\Plainkeywords{Dependent Data, Generalized Linear Model, Minorization-Maximization, Quasi-Newton, Social Network, Spillover}

  \Address{
  Cornelius Fritz\\
  School of Computer Science and Statistics\\
  Trinity College Dublin\\
  O'Reilly Institute\\
  Dublin 2, Republic of Ireland \\
  E-mail: \email{c.fritz@tcd.ie}\\
  \\
  Michael Schweinberger\\
  Department of Statistics\\
  The Pennsylvania State University\\
  326 Thomas Building\\
  University Park, PA 16802\\
  E-mail: \email{mus47@psu.edu}
}

\usepackage{tikz}

\newcommand{\bY}{\mathbf{Y}}

\newcommand{\by}{\mathbf{y}}

\newcommand{\mP}{\mathscr{P}}

\usepackage[tikz]{bclogo}

\newcommand{\bt}{\begin{bclogo}[couleur={rgb:orange,0;yellow,0;white,1},arrondi=0.1,logo=\bcplume,ombre=true]}
	\newcommand{\et}{\end{bclogo}\s}
\newcommand{\btt}{\begin{box}}
	\newcommand{\ett}{\end{box}}

\newcommand{\btheorem}{\begin{bclogo}[couleur={rgb:orange,0;yellow,0;white,1},arrondi=0.1,logo=\bcplume,ombre=true]{Theorem}}
	\newcommand{\ettheorem}{\end{bclogo}}

\newcommand{\bsh}{\begin{bclogo}[couleur={rgb:orange,0;yellow,0;white,1},arrondi=0.1,logo=\bcpanchant,ombre=true]}
	\newcommand{\esh}{\end{bclogo}}

\newcommand{\benum}{\begin{enumerate}}
	\newcommand{\eenum}{\end{enumerate}}

\newcommand{\bq}{\begin{quote}\em}
	\newcommand{\eq}{\end{quote}}
\newcommand{\bbq}{\begin{quote}\bf\em}
	\newcommand{\ebq}{\end{quote}}

\newcommand{\ind}{\msim\limits^{\mbox{\tiny ind}}}

\renewcommand{\bar}{\overline}

\newcommand{\mR}{\mathbb{R}}

\newcommand{\mbE}{\mathbb{E}}

\newcommand{\mbP}{\mathbb{P}}

\newcommand{\ghost}[1]{}
\newcommand{\ba}{\begin{array}{llllllllll}}
	\newcommand{\ea}{\end{array}}
\newcommand{\bea}{\begin{equation}\begin{array}{llllllllll}}
		\newcommand{\eea}{\end{array}\end{equation}}
\newcommand{\nat}{\btheta}
\newcommand{\Nat}{\bTheta}
\newcommand{\truth}{\nat^\star}

\newcommand{\beno}{\begin{equation}\begin{array}{llllllllll}\nonumber}
		\newcommand{\be}{\begin{equation}\begin{array}{llllllllll}}
				\newcommand{\ee}{\end{array}\end{equation}}
		\newcommand{\bi}{\begin{itemize}}
			\newcommand{\ei}{\end{itemize}}
		\newcommand{\ben}{\begin{enumerate}}
			\newcommand{\een}{\end{enumerate}}

		\newcommand{\dsum}{\displaystyle\sum\limits}
		\newcommand{\dint}{\displaystyle\int\limits}
		\newcommand{\dprod}{\displaystyle\prod\limits}
		\newcommand{\dd}{\mathop{\mbox{d}}\nolimits}

		\newcommand{\mN}{\mathscr{N}}

		\newcommand{\mV}{\mathscr{V}}
		
		\newcommand{\s}{\vspace{0.25cm}}
		
		\newcommand{\bx}{\bm{x}}

		\newcommand{\bX}{\bm{X}}
		\newcommand{\mX}{\mathscr{X}}

		\newcommand{\mY}{\mathscr{Y}}
		
		\newcommand{\mZ}{\mathscr{Z}}
		\newcommand{\mD}{\mathscr{D}}

		\newcommand{\bz}{\bm{z}}
		\newcommand{\bZ}{\bm{Z}}

		\newcommand{\bTheta}{\bm\Theta}
		
		\newcommand{\btheta}{\boldsymbol{\theta}}

		\newcommand{\mbV}{\mathbb{V}}

		\newcommand{\msim}{\mathop{\rm \sim}}
		
		\newcounter{counterexample}

		\newcounter{definition}
		
		\newcounter{theorem}

		\newcounter{proposition}

		\newcounter{result}

		\newcounter{tproof}
		\newcounter{corollary}

		\newcounter{cproof}
		\newcounter{lemma}

		\newcounter{com}

		\newcounter{lproof}
		\newcounter{assumption}

		\newif\ifmydraft
		\mydraftfalse

		\ifmydraft
		
		\pagecolor{black!20}
		\else
		
		\fi

		\ifmydraft
		
		\pagecolor{black!20}
		\fi
		
		\usepackage{multicol}

		\makeatletter
		\newcommand*{\deq}{\mathrel{\rlap{%
					\raisebox{0.3ex}{$\m@th\cdot$}}%
				\raisebox{-0.3ex}{$\m@th\cdot$}}=}
		\makeatother

\definecolor{spdred}{RGB}{227,0,15}

\usepackage{fontawesome5}
\begin{document}

\section{Regression under Interference} 
\label{sec:intro}

Generalized linear models (GLMs) are widely used in data science,
helping study relationships among binary,
count-,
and real-valued predictors $X$ and outcomes $Y$ and make model-based predictions.
These regression models,
implemented in the \proglang{R} core function \code{glm()} \citep{cran} and other statistical software packages,
assume that the attributes $(X_i, Y_i)$ of units $i$ are unaffected by the attributes $(X_j, Y_j)$ of all other units $j$ in the population of interest.
Such independence assumptions may be violated in connected populations due to spillover,
contagion,
and other complex phenomena inducing dependence among attributes of connected population members.
For example,
if teenager $i$ is exposed to advertisements of designer clothes on social media ($X_i = 1$),
then $i$ may purchase the advertised designer clothes ($Y_i = 1$).
The effect of $X_i$ on $Y_i$ can spill over to other teenagers for at least two reasons.
First,
teenager $i$ may share the advertisements with friend $j$,
who may purchase the advertised designer clothes ($Y_j = 1$).
Second,
teenager $j$ may observe friend $i$ purchasing the designer clothes ($Y_i = 1$) and may decide to purchase them as well ($Y_j = 1$).
As a consequence,
the outcomes $Y_i$ and $Y_j$ of teenagers $i$ and $j$ are dependent conditional on the exposures $X_i$ and $X_j$ of $i$ and $j$ to advertisements when $i$ and $j$ are connected ($Z_{i,j} = 1$).
Studying relationships among attributes in connected populations requires a regression framework for dependent binary,
count-,
and real-valued attributes $(X_i, Y_i)$ and connections $Z_{i,j}$.

\proglang{R} package \pkg{iglm} offers a unifying regression framework for binary-,
count-,
and real-valued attributes $(X_i, Y_i)$ and connections $Z_{i,j}$,
where attributes of connected units may be dependent due to spillover,
contagion,
and other forms of interference in connected populations \citep{FrScBhHu24}.
The name of the package and its core function \code{iglm()} reflects its purpose by combining \code{i} (interference) with \code{glm()} (generalized linear models).
\proglang{R} package \pkg{iglm} assumes that observations of attributes $(X_i, Y_i)$ and connections $Z_{i,j}$ are available.
Examples of connections include interactions on social media,
friendships and working relationships among employees,
economic transactions among individual or corporate actors,
interactions among animals,
and interactions among neurons \citep[e.g.,][]{Ko17}.

\subsection{Comparison with Other Packages}
\label{sec:comparison}

While no unifying regression framework for dependent binary,
count-,
and real-valued attributes $(X_i, Y_i)$ and connections $Z_{i,j}$ exists,
there are three classes of conditional and joint probability models for dependent attributes and connections:
\bi 
\item {\bf Conditional models for dependent attributes given connections} include autologistic actor attribute models implemented in packages \pkg{PNet} and \pkg{ALAAMEE} \citep{pnet,stivala2024},
which are limited to binary attributes.
\item {\bf Conditional models for dependent connections given attributes} include additive and multiplicative effects models \citep{Ho18} implemented in \proglang{R} package \pkg{amen} \citep{amen},
and exponential random graph models implemented in \pkg{ergm} \citep{ergm2023} and \pkg{Bergm} \citep{bergm.jss}.
\item {\bf Joint probability models} for dependent,
categorical or continuous attributes and connections include continuous-time Markov processes \citep{SnStSc05,NiSn17} implemented in \proglang{R} package \pkg{RSiena} \citep{rsiena2025};
exponential random network models for categorical or continuous attributes and connections \citep{FeHa12} implemented in \pkg{ernm} \citep{ernm2025};
and multivariate normal models for continuous attributes and connections \citep{FoHo15},
which are not implemented in publicly available software.
\ei 

The closest \proglang{R} packages are \pkg{RSiena} and \pkg{ernm}.
\proglang{R} package \pkg{RSiena} requires temporal data,
whereas \pkg{iglm} does not.
\proglang{R} packages \pkg{iglm} and \pkg{ernm} reside on the same exponential-family platform \citep{Su19},
but differ in purpose,
models,
methods,
and theoretical guarantees for statistical estimators.
The purpose of \proglang{R} package \pkg{iglm} is to offer data scientists a unifying regression framework for dependent binary,
count-,
and real-valued attributes in connected populations with interference,
whereas the purpose of \proglang{R} package \pkg{ernm} is to extend exponential-family random graph models from fixed to random predictors.
Some (but not all) of the models implemented in \proglang{R} package \pkg{ernm} exhibit model near-degeneracy,
as demonstrated by \citet{FeHa12}.
Potential remedies for model near-degeneracy---including geometrically weighted model terms \citep{SnPaRoHa04},
regularized homophily terms \citep{FeHa12},
and penalization methods \citep{BlHa23}---weaken the main advantages of parametric models:
interpretability and high rates of convergence for statistical estimators.
First,
geometrically weighted model terms are not easy to interpret,
and regularized homophily terms are not coordinatewise affine functions of attributes and therefore cannot be interpreted as regression terms when attributes are count- or real-valued \citep[cf.\ Proposition 1,][]{FrScBhHu24}.
Second,
the theoretical properties of the approximate maximum likelihood estimators implemented in \proglang{R} package \pkg{ernm} are unknown,
but it is known that model near-degeneracy of network models can depress the rate of convergence of unpenalized maximum likelihood estimators \citep[cf.\ the discussion of Theorem 1 in][]{StSc20},
whereas penalization can alleviate model near-degeneracy but may induce bias.
Worse,
the mentioned remedies alleviate symptoms of model near-degeneracy without addressing its roots:
the unrealistic homogeneity and long-range dependence assumptions of the edge-and-triangle model and related models,
which induce strong dependence in large populations \citep{St86,Jo99}.
By contrast,
\proglang{R} package \pkg{iglm} addresses the roots of model near-degeneracy by capturing unobserved heterogeneity in the propensities of population members to form connections,
and by leveraging additional structure in the form of overlapping neighborhoods to allow the dependence among attributes $(X_i, Y_i)$ and connections $Z_{i,j}$ to propagate throughout the population,
while controlling the amount of dependence induced by the model.
By controlling dependence,
additional structure facilitates provable theoretical guarantees for statistical estimators,
including consistency results and rates of convergence for maximum pseudo-likelihood estimators \citep[cf.\ Theorem 1 and Corollary 1,][]{FrScBhHu24}.

\subsection{Advantages of R Package iglm}
\label{sec:adv}

Compared to existing packages,
\proglang{R} package \pkg{iglm} offers the following benefits:
\begin{itemize}
\item {\bf Unifying regression framework for attributes $(X_i, Y_i)$ and connections $Z_{i,j}$:}\break 
\proglang{R} package \pkg{iglm} offers data scientists a unifying regression framework for binary,
count-,
and real-valued attributes $(X_i, Y_i)$ and connections $Z_{i,j}$.
These regression models allow attributes $(X_i, Y_i)$ and connections $Z_{i,j}$ to be dependent,
enabling the study of spillover,
contagion,
and other complex phenomena in connected populations.
\item {\bf Joint, conditional, and marginal probability models:}
\proglang{R} package \pkg{iglm} focuses on regression models based on joint probability models for dependent attributes $(X_i, Y_i)$ and connections $Z_{i,j}$,
but it allows users to build conditional and marginal models by fixing or omitting predictors $X_i$ or connections $Z_{i,j}$.
\item {\bf Interpretable regression models:}
The regression framework implemented in \proglang{R} package \pkg{iglm} can be viewed as an extension of GLMs from independent attributes $(X_i, Y_i)$ to dependent attributes $(X_i, Y_i)$ and connections $Z_{i,j}$.
As a consequence,
users can interpret results along the lines of GLMs,
including logistic regression,
Poisson regression,
and linear regression models.
\item {\bf Unobserved heterogeneity:}
\proglang{R} package \pkg{iglm} allows users to capture unobserved heterogeneity in the propensities of units to connect with others by using degree terms,
in addition to capturing observed heterogeneity via covariates.
\item {\bf Local dependence in small and large populations:}
\proglang{R} package \pkg{iglm} leverages additional structure in the form of overlapping neighborhoods,
which allows the dependence among attributes $(X_i, Y_i)$ and connections $Z_{i,j}$ to propagate throughout the population.
Neighborhood structure respects the local nature of small and large populations while controlling dependence,
facilitating provable theoretical guarantees for statistical estimators.
\proglang{R} package \pkg{iglm} supplies an extensive list of model terms with local dependence among attributes $(X_i, Y_i)$ and connections $Z_{i,j}$.
\item {\bf Extendability:}
Users can contribute custom-built model terms to \proglang{R} package \pkg{iglm},
making it a versatile platform for regression under interference in connected populations.
\item {\bf Scalability:}
\proglang{R} package \pkg{iglm} can estimate models from small and large populations by solving a convex optimization problem based on pseudo-likelihoods using Minorization-Maximization and Quasi-Newton methods.
The estimation algorithm does not require Markov chain Monte Carlo simulations,
although uncertainty quantification does rely on simulations.
The computing time of the estimation algorithm is of order $O(N^2\, q^2 + q^3)$,
assuming that the population consists of $N$ units and the model includes $O(N)$ degree terms along with $q$ other model terms.
These methods enable applications with up to $N = $ 10,000 units and large-scale simulation studies with up to 4,000 units,
as demonstrated by \citet{FrScBhHu24}.
In applications,
the computing time will depend on the data (e.g., how sparse the network and neighborhood structure are) and the model (e.g., the number and functional form of model terms).
\end{itemize}
The regression framework,
its scalable methods,
and its theoretical guarantees for statistical estimators are facilitated by additional structure in the form of neighborhoods.
We first explain how additional structure facilitates them,
demonstrate in Section \ref{sec:misspecification} that statistical estimators are robust against small-to-moderate misspecifications of neighborhoods,
and discuss in Section \ref{sec:discussion} how to learn neighborhoods from data.
Throughout the remainder of the manuscript,
we write $\bX = (X_i)$,
$\bY = (Y_i)$,
and $\bZ = (Z_{i,j})$.

\s

{\bf Remark: Additional Structure}
The idea of additional structure in the form of neighborhoods was inspired by the observation that the model near-degeneracy of some (but not all) exponential-family random graph models implemented in \proglang{R} package \pkg{ergm} is rooted in unrealistic homogeneity and long-range dependence assumptions stemming from a lack of structure.
\citet{St86} and \citet{Jo99} point out that the edge-and-triangle model implemented in \proglang{R} package \pkg{ergm} lacks structure and allows each edge to depend on a growing number of other edges,
which can lead to model near-degeneracy \citep{Ha03p,Sc09b,ChDi11}.
In other words,
the probability mass of sufficient statistics $s(\bZ)$ (e.g., the number of edges and triangles) may not be concentrated around the corresponding means $\mbE\, s(\bZ)$,
but may be concentrated near the boundary of the mean-value parameter space of the exponential family (e.g., graphs with almost all possible edges and triangles).
To address model near-degeneracy,
\citet{ScHa13} recommend leveraging additional structure,
which can be observed or unobserved,
and use the additional structure to build models with local dependence.
Examples of additional structure are non-overlapping subpopulations \citep{ScSt19},
overlapping subpopulations \citep{StSc20},
and overlapping neighborhoods \citep{FrScBhHu24}.
Additional structure in the form of overlapping neighborhoods allows the dependence among attributes $(X_i, Y_i)$ and connections $Z_{i,j}$ to propagate throughout the population
while guaranteeing that the dependence is not too strong,
provided that the neighborhoods are not too large and do not overlap too much.
By controlling dependence,
additional structure offers the following advantages:
\bi 
\item[1.] If dependence is controlled,
the probability mass of any smooth function
$f(\bX, \bY, \bZ)$ concentrates around its mean $\mbE\, f(\bX, \bY, \bZ)$,
which can be made formal by using concentration inequalities for Lipschitz functions of dependent random variables \citep[e.g.,][]{ScSt19,StSc20}.
If the mean $\mbE\, f(\bX, \bY, \bZ)$ is in the interior of the mean-value parameter space of the exponential family \citep{Br86},
the bulk of the probability mass of $f(\bX, \bY, \bZ)$ is concentrated in the interior of the mean-value parameter space rather than its boundary.
In other words,
additional structure helps solve the model near-degeneracy problem.
\item[2.] Since additional structure helps control dependence,
users can choose simple and interpretable model terms without facing model near-degeneracy.
Among them are model terms that are coordinatewise affine functions of attributes $(X_i, Y_i)$,
which implies that the conditional densities of binary, 
count-,
and real-valued attributes $(X_i, Y_i)$ can be represented by GLMs \citep[Proposition 1,][]{FrScBhHu24}.
As a consequence,
the statistical framework can be interpreted as a regression framework for dependent attributes $(X_i, Y_i)$ and connections $Z_{i,j}$.
\item[3.] The regression framework implies that the conditional densities of attributes $(X_i, Y_i)$ are univariate exponential-family densities with known normalizing constants (e.g.,
Bernoulli,
Poisson,
or Normal),
enabling scalable inference based on pseudo-likelihoods.
\item[4.] By controlling dependence,
provable theoretical guarantees can be obtained for statistical estimators,
including consistency \citep{ScSt19},
rates of convergence \citep{StSc20},
minimax results \citep{St24},
and Berry-Esseen-type bounds for normal approximations \citep{St24}.
\ei
The first publication to leverage additional structure for building a regression framework for dependent attributes $(X_i, Y_i)$ and connections $Z_{i,j}$ with provable theoretical guarantees for statistical estimators is \citet{FrScBhHu24}, 
which is implemented in \proglang{R} package \pkg{iglm}.

\subsection{Architecture of R Package iglm}

\proglang{R} package \pkg{iglm} is a free,
publicly available,
and platform-independent \proglang{R} package that can be downloaded from the website \url{https://cran.r-project.org} of the Comprehensive R Archive Network (CRAN) \citep{cran}.
It can be installed and loaded in \proglang{R} as follows:
\begin{verbatim}
R> install.packages("iglm")
R> library(iglm)
\end{verbatim}
\proglang{R} package \pkg{iglm} implements an object-oriented architecture via \code{R6} classes \citep{R6} instead of functional object-oriented systems in \proglang{R},
such as \code{S3}.
In a functional class system,
methods belong to generic functions rather than classes,
and dispatch depends on class attributes of arguments.
Updating objects relies on copy-on-modify semantics, 
accruing computational overhead \citep{wickham2019}. 
By contrast,
\code{R6} classes come with the following benefits:
\begin{itemize}
\item \textbf{Unified objects and methods:} 
The \code{R6} classes \code{iglm} and \code{iglm.data} bundle all \code{R} functions belonging to a class in so-called methods. 
For example, 
upon constructing an \code{iglm} object called \code{m}, 
one can use \code{R} functions \code{m$estimate()}, 
\code{m$simulate()}, 
and \code{m$assess()} to estimate, 
simulate, 
and assess model \code{m},
respectively.
This syntax is aligned with the class system in other popular software languages such as \proglang{Python} or \proglang{C++}, 
making \pkg{iglm} more accessible to data scientists who are less familiar with \code{R}.
\item \textbf{Changes of objects without memory duplication:} 
Objects of \pkg{R6} classes are environments in \proglang{R},
enabling changes of objects via direct reference and eliminating the need to create copies of objects.
\item \textbf{Design consistency:} 
The design of the \code{R6} class \code{iglm} places a strong emphasis on consistency between nested classes. 
For example, 
all results are saved in the nested \code{R6} class \code{results},
which is consistent with the main object \code{iglm}: 
e.g.,
the command \code{m$results$plot(trace = TRUE)} provides trace plots of all estimates and if the model is updated,
estimates and model-based predictions under different models will be deleted. 
\item \textbf{Natural data sharing:} 
\proglang{R} package \pkg{iglm} allows users to share data across multiple stages of data analysis.
For example,
simulated data are used in two stages of data analysis---to quantify the uncertainty about estimators and to assess models based on simulated data---and \proglang{R} package \pkg{iglm} shares samples across these two stages of analysis.
\end{itemize}

In addition to its dependency on \proglang{R} package \pkg{R6} \citep{R6}, 
\pkg{iglm} depends on \pkg{coda} \citep{coda} to save statistics;
\pkg{igraph} \citep{igraph} to plot data;
and \pkg{Rcpp} \citep{Rcpp} and \pkg{RcppArmadillo} \citep{RcppArmadillo} to speed up computing by outsourcing time-consuming tasks to \proglang{C++}.
All results are based on \proglang{R} package \pkg{iglm} version 1.3 and \proglang{R} version 4.6.1, 
using a MacBook Pro with macOS Tahoe 26.6.2. 
Since pseudo-random number generators can vary across operating systems and \proglang{R} versions, 
running the replication script on other operating systems or \proglang{R} versions can result in small differences.

\subsection{Structure}

We describe the data required by \proglang{R} package \pkg{iglm} in Section \ref{sec:data} and review models, 
methods, 
and implementation in Section \ref{sec:model},
using data on hate speech on the social media platform X with binary outcomes $Y_i \in \{0, 1\}$ and connections $Z_{i,j} \in \{0, 1\}$ as a running example.
Advanced topics are discussed in Section \ref{sec:extensions},
including model comparison and custom-built model terms.
An additional application to real-valued outcomes $Y_i \in \mR$ and connections $Z_{i,j} \in \{0, 1\}$ is presented in Section \ref{sec:application}.

\section{Data} 
\label{sec:data}

\proglang{R} package \pkg{iglm} stores data on $N \geq 2$ units in data object \code{iglm.data}:
\begin{itemize}
    \item \textbf{Predictors $\bX = (X_i)$}: 
    a vector \code{x_attribute} of length $N$, 
    storing unit-level predictors. 
    The data type is specified by the argument \code{family_x} of \code{iglm.data}, 
    which can be binary ($X_i \in \{0, 1\}$: \code{family_x = "binomial"}), 
    count-valued ($X_i \in \{0, 1, \ldots\}$: \code{family_x = "poisson"}), 
    or real-valued ($X_i \in \mathbb{R}$: \code{family_x = "normal"}). 
    The predictors $X_i$ are random:
    e.g.,
    $X_i$ may be the treatment assignment of unit $i$ in a randomized experiment.
    At the time of writing, 
    \proglang{R} package \pkg{iglm} supports a single random predictor $X_i$,
    but an arbitrary number of non-random predictors can be added to the local \proglang{R} environment,
    including unit-dependent predictors (e.g., the demographic background of units $i$) or dyad-dependent predictors (e.g., indicators of whether pairs of units $i$ and $j$ have a shared demographic background).
    Since non-random predictors are viewed as exogenous,
    \proglang{R} package \pkg{iglm} does not store them in data object \code{iglm.data}.
    \item \textbf{Outcomes $\bY = (Y_i)$}: 
    a vector \code{y_attribute} of length $N$, 
    storing unit-level outcomes. 
    The data type is specified by the argument \code{family_y} of \code{iglm.data}, 
    which can be binary ($Y_i \in \{0, 1\}$: \code{family_y = "binomial"}), 
    count-valued ($Y_i \in \{0, 1, \ldots\}$: \code{family_y = "poisson"}), 
    or real-valued ($Y_i \in \mathbb{R}$: \code{family_y = "normal"}). 
    \item \textbf{Connections $\bZ = (Z_{i,j})$}: 
    A connection between units $i$ and $j$ is indicated by $Z_{i,j} \in \{0, 1\}$, 
    where $Z_{i,j} \coloneqq 1$ indicates that units $i$ and $j$ are connected and $Z_{i,j} \coloneqq 0$ otherwise.
    The connections $Z_{i,j}$ can be directed or undirected, in which case $Z_{i,j} = Z_{j,i}$ for all pairs of units $i$ and $j$.
    Self-connections are excluded by setting $Z_{i,i} \coloneqq 0$ for all units $i$.
    The directed or undirected connections can be represented by an $N \times N$-adjacency matrix $\bZ \coloneqq (Z_{i,j}) \in \{0,1\}^{N \times N}$ or an $M \times 2$-edge matrix $\bm{E} \coloneqq (E_{m,k}) \in \{1, \ldots, N\}^{M \times 2}$, 
    where $M \geq 1$ is the number of observed connections and $E_{m,1} = i$ and $E_{m,2} = j$ indicate that the $m$th observed connection is between units $i$ and $j$. 
    In the case of undirected connections,
    the $M \times 2$-edge matrix lists all pairs of units $i$ and $j$ with $i < j$ such that $Z_{i,j} = 1$,
    and the $N \times N$-adjacency matrix $\bZ$ is symmetric.
    If $N$ is large and the network is sparse,
    edge lists require less memory than adjacency matrices and are preferable. 
    \item \textbf{Neighborhoods $\mN_1, \ldots, \mN_N$:}
    If additional structure in the form of neighborhoods $\mN_1, \ldots, \mN_N$ of units $1, \ldots, N$ is available, 
    it can be imported by specifying the option \code{neighborhood};
    note that these neighborhoods are viewed as exogenous to the network $\bZ$.
    The neighborhood structure can be provided in the same form as the network $\bZ$,
    i.e., 
    either in the form of an $N \times N$-adjacency matrix or as an $L \times 2$-edge matrix specifying which pairs of units $i$ and $j$ are neighbors,
    where $L$ denotes the total number of pairs of neighbors.
    The neighborhoods $\mN_i$ do not have to satisfy any constraints as long as $\mN_i \subseteq \mP$,
    where $i \in \mN_i$ and $\mN_i = \mP$ are permissible.
    If there is partial information about who is a neighbor of whom,
    supersets of neighborhoods including all suspected neighbors can be specified.
    We demonstrate in Section \ref{sec:misspecification} that statistical estimators are robust against small-to-moderate misspecifications of neighborhoods.
    If the neighborhoods are unknown and cannot be covered by supersets,
    users can learn them from data.
    We discuss in Section \ref{sec:discussion} how neighborhoods can be learned from data.
    If no neighborhoods are provided,
    it is assumed that $\mN_i = \mP$ for all $i \in \mP$.
    \end{itemize}
Throughout the remainder of the manuscript,
$\mathbb{I}(\cdot)$ is an indicator function,
which is $1$ if its argument is true and is $0$ otherwise,
and 
\beno 
c_{i,j} 
&\coloneqq& \mathbb{I}(\mN_i \cap \mN_j \neq \emptyset)
\ee
is an indicator of whether the neighborhoods $\mN_i$ and $\mN_j$ of units $i$ and $j$ overlap.

\paragraph*{Example: Hate speech on X}
As a running example,
we use data collected by \citet{kim_attention_2022}.
The data concern hate speech by $N = 495$ U.S.\ state legislators on the social media platform X in the six months preceding the January 6, 2021 insurrection at the U.S.~Capitol.  
We use a subset of the data,
consisting of legislators in California (CA), 
Florida (FL), 
Illinois (IL), 
New York (NY), 
Pennsylvania (PA),
and Texas (TX).
The outcome $Y_i \in \{0, 1\}$ indicates whether posts of legislator $i$ were identified as hate speech by Large Language Models \citep{FrScBhHu24},
where $Y_i \coloneqq 1$ if one or more posts by legislator $i$ were classified as hate speech and $Y_i \coloneqq 0$ otherwise.
The predictor $X_i \in \{0, 1\}$ is $X_i \coloneqq 1$ if legislator $i$ is Republican and $X_i \coloneqq 0$ otherwise.  
Other predictors include gender ($v_{i,1} \coloneqq 1$ if $i$ is female and $v_{i,1} \coloneqq 0$ otherwise),
race ($v_{i,2} \coloneqq 1$ if $i$ is white and $v_{i,2} \coloneqq 0$ otherwise), 
and state ($v_{i,3} \in \{\mbox{CA, FL, IL, NY, PA, TX}\}$).
On the social media platform X, 
users can mention or repost posts by others.
We set $Z_{i,j} \coloneqq 1$ if legislator $i$ mentioned or reposted posts by legislator $j$ between January 6, 2020 and January 6, 2021 and set $Z_{i,j} \coloneqq 0$ otherwise.
The connections $Z_{i,j}$ are directed,
so $Z_{i,j}$ may not equal $Z_{j,i}$.
In line with \citet{FrScBhHu24}, 
the neighborhood $\mN_i$ of legislator $i$ is the union of the set of other legislators that legislator $i$ follows on X and $i$ itself. 

If the vectors \code{x_attribute} and \code{y_attribute} and the matrices \code{z_network} and \code{neighborhood} are available in the local \proglang{R} environment, 
we can instantiate data object \code{iglm.data} as follows:
\begin{verbatim}
R> data.object <- iglm.data(x_attribute = x_attribute,
+                           family_x = "binomial",
+                           y_attribute = y_attribute,
+                           family_y = "binomial",
+                           z_network = z_network,
+                           directed = TRUE,
+                           neighborhood = neighborhood)
\end{verbatim}
The family of connections $Z_{i,j}$ does not need to be specified,
because \proglang{R} package \pkg{iglm} assumes that all connections are binary,
i.e.,
$Z_{i,j} \in \{0, 1\}$.
As mentioned,
the non-random predictors $v_{i,1}$, 
$v_{i,2}$, 
and $v_{i,3}$ are viewed as exogenous and are not stored in \code{data.object},
but these predictors can be added to the local \proglang{R} environment and can be used in models.

\paragraph*{Descriptive Statistics}

\begin{figure}[t!]
\centering
\includegraphics[width=0.45\textwidth]{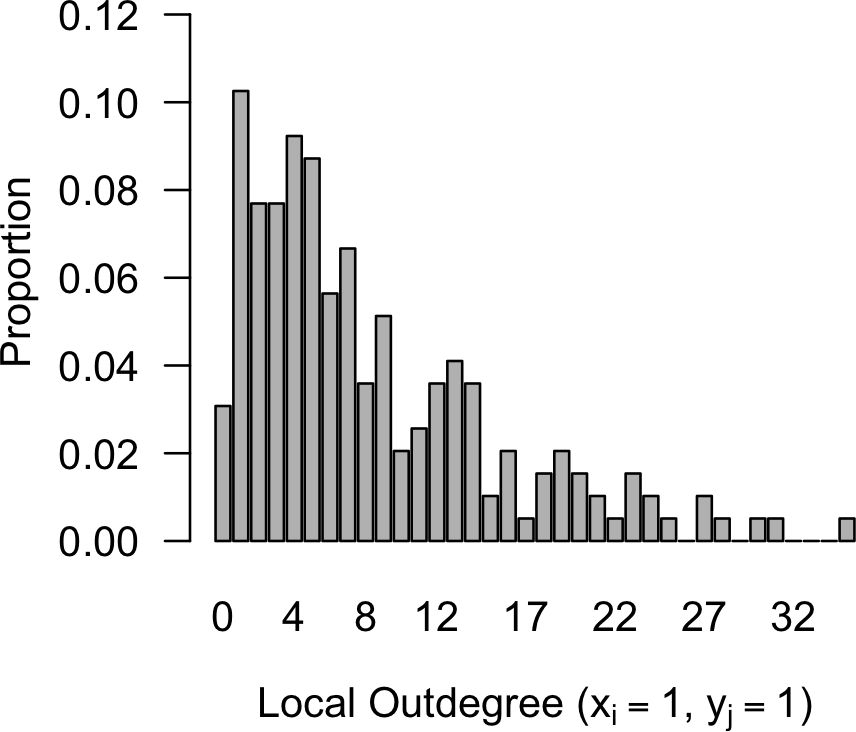}
\includegraphics[width=0.45\textwidth]{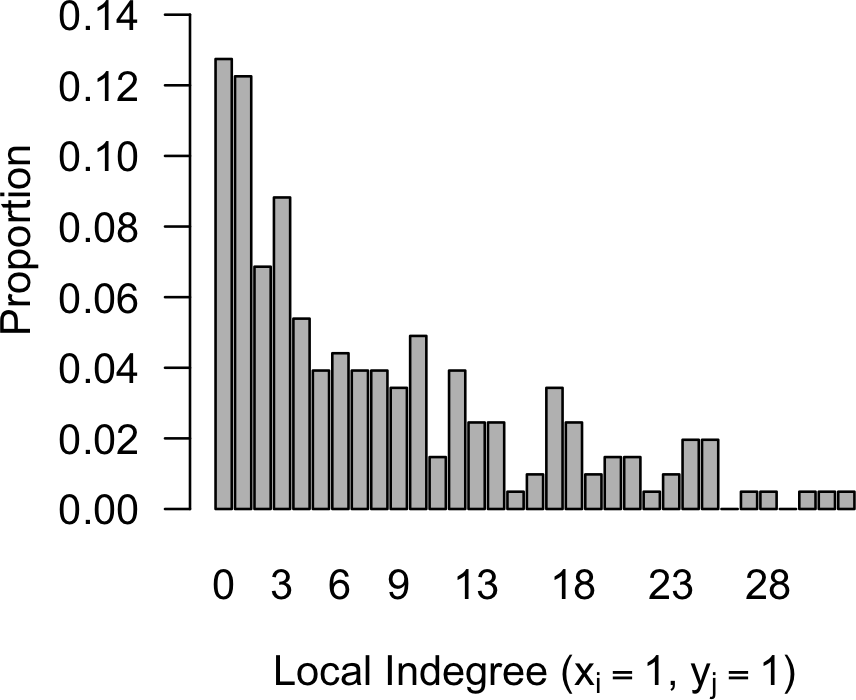}
\caption{
\label{fig:spillover_deg}
Hate speech on X:
Observed in- and outdegrees of U.S.\ state legislators in the subnetwork consisting of connections among pairs of legislators $\{i, j\}$ with $c_{i,j}\, x_i\, y_j = 1$ or $c_{i,j}\, x_j\, y_i = 1$.
Such connections represent the channels that enable spillover and therefore the in- and outdegrees in the described subnetwork are called spillover in- and outdegrees.
}
\end{figure}

Printing \code{data.object} provides descriptive statistics, 
including the number of units ($N = 495$),
an indicator of whether the network is directed,
the number of connected pairs of units ($\sum_{(i,j) \in \mD} Z_{i,j} = $ 9,218,
out of which 9,135 are connections between legislators with overlapping neighborhoods),
the number of local dyads---i.e.,
pairs of units with overlapping neighborhoods ($\sum_{(i,j) \in \mD} c_{i,j} = $ 60,912, corresponding to 24.9$\%$ of all dyads),
and summaries of attributes $(X_i, Y_i)$:
\begin{verbatim}
R> data.object 
iglm.data object
  units                       : 495
  directed                    : TRUE
  connections (fixed = FALSE) : 9218
  local dyads                 : 60912 / 244530 (24.9%)
  
Attribute summaries
  x_attribute (fixed = FALSE) : binomial mean = 0.394
  y_attribute                 : binomial mean = 0.412
\end{verbatim}
Users can explore the data using the following methods supplied by data object \code{iglm.data}:
\begin{itemize}
\item \code{x_dist()} and \code{y_dist()} plot the empirical distributions of attributes $X_i$ and $Y_i$,
respectively.
\item \code{deg_dist(plot, x_i, x_j, y_i, y_j, mode)} computes and plots the degrees of units,
i.e.,
the numbers of connections of units.
If connections are directed,
the function computes and plots both the distribution of indegrees (the numbers of incoming connections of units) and the distribution of outdegrees (the numbers of outgoing connections of units).
The function can compute degrees restricted to connections between units with specified attributes
by specifying the arguments \code{x_i, x_j, y_i, y_j}.
The connections can be restricted to units with overlapping neighborhoods ($c_{i,j} = 1$) by setting \code{mode = "local"}.
If attributes are binary,
users can plot the local spillover in- and outdegree distributions in the subnetwork consisting of connections among pairs of units $\{i, j\}$ with $c_{i,j}\, x_i\, y_j = 1$ or $c_{i,j}\, x_j\, y_i = 1$ as follows:
\begin{verbatim}
R> data.object$deg_dist(plot = TRUE, x_i = 1, y_j = 1, mode = "local")    
\end{verbatim}
The resulting plots are shown in Figure~\ref{fig:spillover_deg}.
For non-binary attributes, 
the spillover in- and outdegree distributions are computed based on binarized attributes $\widetilde{x}_i \coloneqq \mathbb{I}(x_i > \bar{x})$ and $\widetilde{y}_i \coloneqq \mathbb{I}(y_i > \bar{y})$, 
where $\bar{x}$ and $\bar{y}$ are sample means.
\item \code{geo_dist(plot, x_i, x_j, y_i, y_j, mode)} computes and plots geodesic distances,
i.e.,
the numbers of pairs of units with shortest paths of length $1, 2, \ldots$
The arguments \code{x_i, x_j, y_i, y_j} can be used to compute geodesic distances of pairs of units with specified attributes.
Setting \code{mode = "local"} calculates pairwise geodesic distances of units with overlapping neighborhoods ($c_{i,j} = 1$), 
while \code{mode = "global"} computes pairwise geodesic distances of all units ($c_{i,j} \in \{0, 1\}$).
\item \code{dsp_dist(plot, x_i, x_j, y_i, y_j, mode)} computes and plots the distribution of the numbers of pairwise shared partners,
i.e.,
the numbers of pairs of units---either connected or unconnected---with $0, 1, \ldots$ shared partners.  
The arguments \code{x_i, x_j, y_i, y_j} can be used to compute the numbers of pairwise shared partners with specified attributes,
while \code{mode} can be \code{local} ($c_{i,j} = 1$) or \code{global} ($c_{i,j} \in \{0, 1\}$).
\item \code{esp_dist(plot, x_i, x_j, y_i, y_j, mode)} computes and plots the distribution of the numbers of edgewise shared partners,
i.e.,
the numbers of connected pairs of units with $0, 1, \ldots$ shared partners. 
The arguments \code{x_i, x_j, y_i, y_j} can be used to compute the numbers of edgewise shared partners with specified attributes,
while \code{mode} can be \code{local} ($c_{i,j} = 1$) or \code{global} ($c_{i,j} \in \{0, 1\}$).
\end{itemize}

\begin{figure}[t!]
  \centering
  \includegraphics[width=\textwidth, trim={140pt 250pt 10pt 200pt}, clip]{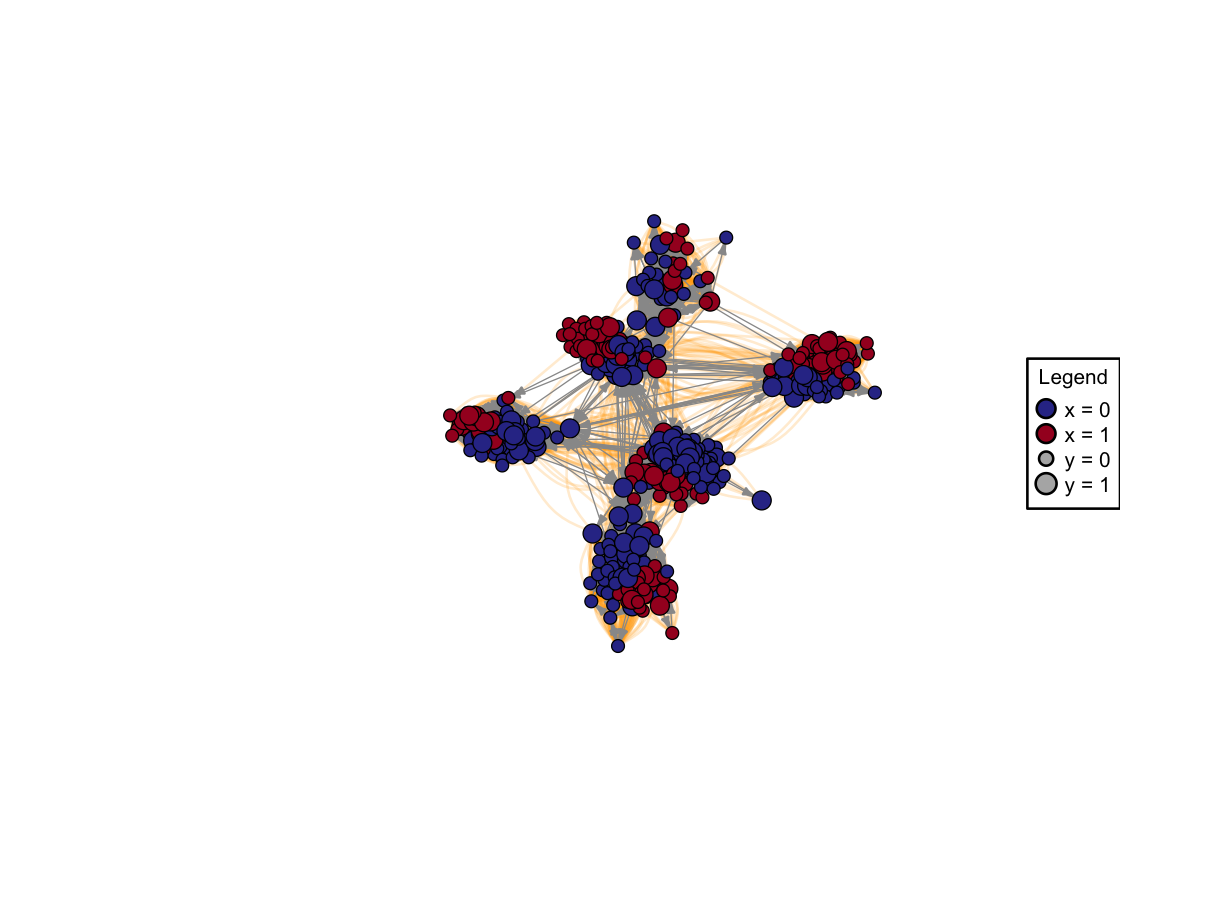}
   \caption{  
   \label{fig:igraph1}
   Hate speech on X: 
   Interactions among U.S.\ state legislators represented by lines between circles. 
   Yellow lines between legislators $i$ and $j$ indicate that the neighborhoods $\mN_i$ and $\mN_j$ of $i$ and $j$ overlap,
   whereas grey lines indicate connections.
   The colors and sizes of circles represent party affiliation $X_i$ and use of hate speech $Y_i$ by legislators $i$. 
   }
\end{figure}

To visualize the data object,
\proglang{R} package \pkg{iglm} relies on tools implemented in \proglang{R} package \pkg{igraph} \citep{igraph}:   
\begin{verbatim}
R> set.seed(123)
R> data.object$plot(edge.width = 0.5, edge.arrow.size = 0.2, 
+                   legend_size = 0.5, legend_size_n_levels = 2)
\end{verbatim}
The resulting plot is shown in Figure \ref{fig:igraph1}.

\section{Regression Framework} 
\label{sec:model}

We describe the regression framework for dependent binary,
count-,
and real-valued attributes $(\bX, \bY)$ and connections $\bZ$ implemented in \proglang{R} package \pkg{iglm},
including model specification (Section \ref{sec:specification}),
estimation (Section \ref{sec:estimation}),
uncertainty quantification (Section \ref{sec:uq}),
simulation (Section \ref{sec:simulation}),
assessment (Section \ref{sec:assessment}),
interpretation (Section \ref{sec:interpretation}),
and robustness of estimators to misspecifications of neighborhoods (Section \ref{sec:misspecification}).

The regression framework is based on probability measures $\{\mbP_{\nat},\, \nat \in \Nat\}$ dominated by a $\sigma$-finite measure $\nu_{\mX,\mY,\mZ}$ with densities $f_{\nat} \coloneqq \dd \mbP_{\nat} / \dd \nu_{\mX,\mY,\mZ}$ of the form
\beno
f_{\nat}(\bx, \by, \bz; \mN) 
&=& \dfrac{1}{c(\nat; \mN)} \left[\, \dprod_{i \in \mathscr{P}} a_{\mX}(x_i)\, a_{\mY}(y_i)\, \exp(\nat_g^\top g_i(x_i^\star, y_i^\star))\right]\s
\\
&\times& \left[\, \dprod_{(i,j) \in \mathscr{D}} a_{\mZ}(z_{i,j})\, \exp(\nat_h^\top h_{i,j}(x_i^\star, x_j^\star, y_i^\star, y^\star_j, \bz; \mN))\right],
\ee 
where 
\begin{itemize}
\item $\mathscr{P} \coloneqq \{1, \ldots, N\}$ is the set of $N \geq 2$ units,
$\mN \coloneqq (\mN_i)_{i \in \mP}$ is the set of neighborhoods $\mN_i \subseteq \mP$ of units $i \in \mP$,
and $\mathscr{D} \subseteq \mP^2$ is the set of pairs of units with possible connections:
\beno 
\mathscr{D} &\coloneqq& 
\begin{cases} 
\{(i,j) \in \mP^2: 1 \le i,\, j \le N, \, i \neq j\} & \text{if connections are directed}\s 
\\
\{(i,j) \in \mP^2: 1 \le i < j \le N\} & \text{if connections are undirected.}
\end{cases}
\label{def:D}
\ee
\item $a_{\mX}(\cdot)$, 
$a_{\mY}(\cdot)$, 
and $a_{\mZ}(\cdot)$ are reference measures that determine the conditional distributions of $X_i$,
$Y_i$, 
and $Z_{i,j}$,
as demonstrated in Section \ref{sec:specification}. 
\item $x_i^\star \coloneqq x_i/\psi_\mathscr{X}$ and $y_i^\star \coloneqq y_i/\psi_\mathscr{Y}$,
where $\psi_\mathscr{X} > 0$ and $\psi_\mathscr{Y} > 0$ are scale parameters that resemble the dispersion parameters of GLMs \citep{MpNj83}.
\item $\nu_{\mX,\mY,\mZ}$ is a $\sigma$-finite product measure of the form
\beno
\nu_{\mX,\mY,\mZ}(\bx, \by, \bz)
&\coloneqq& \left[\, \dprod_{i \in \mP} \nu_{\mX}(x_i)\, \nu_{\mY}(y_i)\right] \left[\, \dprod_{(i,j) \in \mD} \nu_{\mZ}(z_{i,j})\right],
\ee
where $\nu_{\mX}$, 
$\nu_{\mY}$, 
and $\nu_{\mZ}$ are $\sigma$-finite measures that depend on the sample spaces of $X_i$,
$Y_i$,
and $Z_{i,j}$.
\proglang{R} package \pkg{iglm} uses counting measure for $X_i, Y_i \in \{0, 1\}$ and $X_i, Y_i \in \{0, 1, \ldots\}$,
Lebesgue measure for $X_i, Y_i \in \mR$,
and counting measure for $Z_{i,j} \in \{0, 1\}$.
\item $g_i(\cdot)$ is a vector of real-valued functions that describe the relationship between predictors $X_i$ and outcomes $Y_i$,
with a corresponding vector of weights $\nat_g$.
\item $h_{i,j}(\cdot)$ is a vector of real-valued functions that specify how the attributes $(X_i, Y_i)$ and $(X_j, Y_j)$ of units $i$ and $j$ depend on each other, 
and how the connection $Z_{i,j}$ depends on other connections,
with a corresponding vector of weights $\nat_h$.
\item $\nat \coloneqq (\nat_g,\, \nat_h) \in \Nat$ is the vector of all weights with values in $\Nat  \coloneqq \{\nat \in \mR^p: c(\nat; \mN) < \infty\}$ ($p \geq 1$),
where the normalizing constant $c(\nat; \mN)$ is defined by
\beno 
c(\nat; \mN) &\coloneqq& \dint_{\mX \times \mY \times \mZ}\,  &\left[\, \dprod_{i \in \mathscr{P}} a_{\mX}(x_i)\, a_{\mY}(y_i)\, \exp(\nat_g^\top g_i(x_i^\star, y_i^\star))\right]\s
\\
& ~ &\,\,\,\,\, \times &\left[\, \dprod_{(i,j) \in \mathscr{D}} a_{\mZ}(z_{i,j})\, \exp(\nat_h^\top h_{i,j}(x_i^\star, x_j^\star, y_i^\star, y^\star_j, \bz; \mN))\right] \dd \nu_{\mX,\mY,\mZ}(\bx, \by, \bz).
\ee
The set $\Nat$,
consisting of all weights $\nat \in \mR^p$ for which the normalizing constant $c(\nat; \mN)$ is finite,
depends on the sample space of $(\bX, \bY, \bZ)$.
Since \proglang{R} package \pkg{iglm} implements a wide range of regression models for binary-,
count-,
and real-valued attributes $(\bX, \bY)$ and connections $\bZ$,
a discussion of all cases is infeasible,
and we refer interested readers to the related literature on mixed graphical models for binary-, 
count-,
and real-valued attributes $\bY$ \citep[e.g.,][]{mixed,ChWiSh15} and count-valued connections $\bZ$ \citep{Kr11}.
\end{itemize}
While the normalizing constant $c(\cdot)$ and the functions $h_{i,j}(\cdot)$ depend on $\mN$,
we suppress the functional dependence of $c(\cdot)$ and $h_{i,j}(\cdot)$ on $\mN$ throughout the remainder of the manuscript.

The regression framework can capture relationships among attributes $(X_i, Y_i)$ of units $i$ via $g_i(\cdot)$ and dependence among attributes $(X_i, Y_i)$ and $(X_j, Y_j)$ of connected pairs of units $i$ and $j$ ($Z_{i,j} = 1$) via $h_{i,j}(\cdot)$.
We demonstrate in Sections \ref{sec:specification}, \ref{sec:interpretation}, and \ref{sec:application} that these models can be viewed as a natural extension of GLMs from independent attributes $(X_i, Y_i)$ to dependent attributes $(X_i, Y_i)$ and connections $Z_{i,j}$.

{\bf Local Dependence in Overlapping Neighborhoods.}
To ensure that regression models for dependent attributes $(X_i, Y_i)$ and connections $Z_{i,j}$ are well-behaved in small and large populations and obtain provable theoretical guarantees for statistical estimators,
we assume that $(\bX, \bY, \bZ)$ satisfies local dependence,
in the sense that
\be
\label{local.dependence.assumption}
h_{i,j}(x^\star_i, x^\star_j, y^\star_i, y^\star_j, \bz) 
&=& h_{i,j}(z_{i,j}, z_{j,i}) \mbox{ for all $(i, j) \in \mD$ such that } \mN_i\, \cap\, \mN_j = \emptyset.
\ee
The fact that $h_{i,j}(\cdot)$ is allowed to be a function of $z_{i,j}$ and $z_{j,i}$ helps capture mutuality or reciprocity among pairs of units with non-overlapping neighborhoods,
i.e., 
the tendency to reciprocate connections in directed networks.

The local dependence assumption in Equation \eqref{local.dependence.assumption} implies the following:
\bi 
\item The dependence among attributes $(X_i, Y_i)$ and connections $Z_{i,j}$ can propagate throughout the population,
because the neighborhoods of units can overlap.
To demonstrate,
consider a population $\mP \coloneq \{1, 2, 3\}$ consisting of $N = 3$ units $1, 2, 3$ with neighborhoods $\mN_1 = \{1\}$,
$\mN_2 = \{1, 2\}$,
$\mN_3 = \{2, 3\}$.
The neighborhoods of units $1$ and $3$ do not overlap,
but the attributes of units $1$ and $3$ are dependent:
e.g.,
if $\bX$ and $\bZ$ are omitted,
then a marginal model of $\bY \in \{0, 1\}^3$ with $\psi_{\mY} \coloneqq 1$ can be specified by
\beno 
g_i(y_i^\star) 
= 0
\;\mbox{ for all }\; i \in \mP
&\mbox{and}&
h_{i,j}(y_i^\star, y_j^\star) 
= c_{i,j}\, y_i\, y_j \;\mbox{ for all }\; (i, j) \in \mD,
\ee
which implies
\beno 
f_\theta(y_1, y_2, y_3)
&=& \dfrac{1}{c(\theta)}\, \exp(\theta\, y_1\, y_2 + \theta\, y_2\, y_3),
& \theta \in \mR.
\ee
If $\theta \neq 0$,
then
\beno 
f_\theta(y_1, y_3)
&=& \dsum_{y_2 = 0}^1 f_\theta(y_1, y_2, y_3)
&=& \dfrac{1}{c(\theta)} + \dfrac{\exp(\theta\, y_1)\, \exp(\theta\, y_3)}{c(\theta)},
\ee
and it can be shown that
\beno
f_\theta(y_1, y_3)
&\neq& f_\theta(y_1)\, f_\theta(y_3).
\ee
Therefore,
the outcomes $Y_1$ and $Y_3$ are dependent,
despite the fact that the neighborhoods of units $1$ and $3$ do not overlap.
In fact,
all outcomes are dependent,
which shows that the dependence among outcomes propagates throughout the population.
\item While overlapping neighborhoods allow the dependence among attributes and connections to propagate throughout the population,
the dependence induced by the model is not too strong as long as the neighborhoods are not too large and do not overlap too much.
We explain in Section \ref{sec:adv} that additional structure,
coupled with local dependence assumptions,
enables desirable probabilistic properties (e.g., 
the absence of model near-degeneracy) and theoretical guarantees for statistical estimators.
\ei 

\subsection{Specification} 
\label{sec:specification}

\begin{figure}[t!]
\centering
\includegraphics[width=0.5\textwidth]{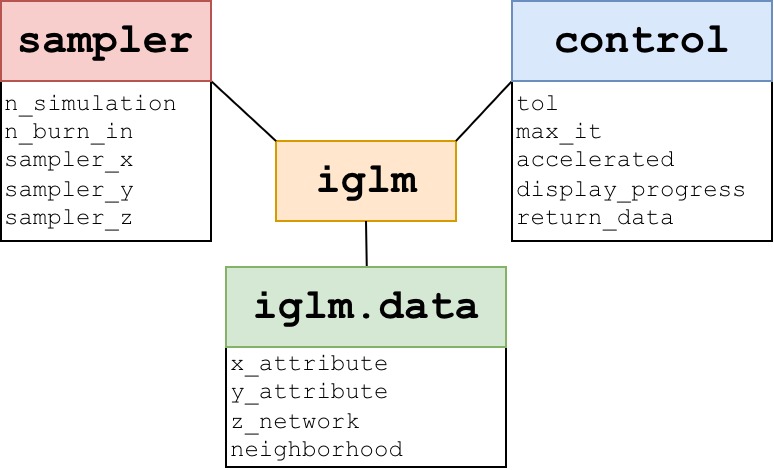}
\caption{
\label{fig:iglm_class}
A graphical representation of the essential parts of class \code{iglm}: objects \code{iglm.data}, 
\code{control}, 
\code{sampler}. 
Additional options supplied by these objects are described in the help function of \proglang{R} package \pkg{iglm}:
see \code{?iglm}. 
}
\end{figure}

The class \code{iglm} supplies statistical tools for data analysis,
including model specification,
estimation,
uncertainty quantification, 
simulation,
and assessment.
A graphical representation of class \code{iglm} can be found in Figure \ref{fig:iglm_class}.
An object of class \code{iglm} is defined via three components:
\bi 
\item \code{formula} specifies the data and model;
\item \code{control} specifies the estimation algorithm described in Section \ref{sec:estimation};
\item \code{sampler} specifies the simulation algorithm described in Section \ref{sec:simulation}. 
\ei

In line with GLMs for independent attributes $(X_i, Y_i)$ implemented in the \proglang{R} core function \code{glm()},
regression models for dependent attributes $(X_i, Y_i)$ and connections $Z_{i,j}$ are specified by an object of type \code{formula}:
\begin{verbatim}
formula = data.object ~ term1 + term2 + ...
\end{verbatim}
The left-hand side specifies the response variable \code{data.object}, 
while the right-hand side specifies the model using terms \code{term1} and \code{term2} separated by the symbol \code{+}\,.
The dots \code{...} represent additional model terms.
In contrast to GLMs for independent attributes $(X_i, Y_i)$ implemented in the \proglang{R} core function \code{glm()},
the attributes $(X_i, Y_i)$ and $(X_j, Y_j)$ of units $i$ and $j$ with connections $Z_{i,j}$ can be dependent,
and connections $Z_{i,j}$ can depend on other connections by including suitable model terms.

\s 

{\bf Left-hand side of \code{formula}: iglm.data}

\s

The class \code{iglm.data} is described in Section \ref{sec:data}. 
We discuss the specification of the \code{family} of predictors $X_i$ and outcomes $Y_i$,
which determines the reference measures $a_{\mX}(\cdot)$ and $a_{\mY}(\cdot)$.

We first consider outcomes $Y_i$ and write $\bY_{-i}$ to denote the set of outcomes $\bY$ excluding $Y_i$:
\begin{itemize}
\item \code{set_family_y("binomial")} sets $a_\mY(y_i) \coloneqq \mathbb{I}(y_i \in \{0,1\})$,
which implies that the conditional distribution of $Y_i \mid (\bX, \bY_{-i}, \bZ) = (\bx, \by_{-i}, \bz)$ is Bernoulli with mean\break $\exp(\eta_{\mY,i})/(1 + \exp(\eta_{\mY,i}))$,
where $\eta_{\mY,i} \in \mR$ is a linear predictor that depends on \code{formula}.
\item \code{set_family_y("poisson")} sets $a_\mY(y_i)
\coloneqq 1/(y_i!) \; \mathbb{I}(y_i \in \{0, 1, \dots\})$,
which implies that the conditional distribution of $Y_i \mid (\bX, \bY_{-i}, \bZ) = (\bx, \by_{-i}, \bz)$ is Poisson with mean $\exp(\eta_{\mY,i})$.
\item \code{set_family_y("normal")} sets 
\beno 
a_\mY(y_i)
&\coloneqq& \dfrac{1}{\sqrt{2\, \pi\, \psi_{\mY}}} \, \exp \left(- \dfrac{y_i^2}{2\, \psi_{\mY}}\right)\, \mathbb{I}(y_i \in \mR),
\ee
which implies that the conditional distribution of $Y_i \mid (\bX, \bY_{-i}, \bZ) = (\bx, \by_{-i}, \bz)$ is Normal with mean $\eta_{\mY,i}$ and variance $\psi_{\mY} > 0$.
\end{itemize}
In all three cases,
the linear predictor $\eta_{\mY,i}$ depends on the joint probability model of $(\bX, \bY, \bZ)$,
which is specified by \code{formula}.
We provide examples of linear predictors $\eta_{\mY,i}$ when $Y_i \in \{0, 1\}$ (Section \ref{sec:interpretation}) and $Y_i \in \mR$ (Section \ref{sec:application}).

The conditional distributions of predictors $X_i$ and outcomes $Y_i$ can be represented by GLMs.
The specific GLMs depend on the sample spaces and reference measures $a_{\mX}(.)$ and $a_{\mY}(.)$ of $X_i$ and $Y_i$. 
The conditional distributions of connections $Z_{i,j} \in \{0, 1\}$ are Bernoulli,
assuming that $a_{\mZ}(z_{i,j}) \coloneqq 1$ if $z_{i,j} \in \{0, 1\}$ and $a_{\mZ}(z_{i,j}) \coloneqq 0$ otherwise,
and can be represented by logistic regression models.

\proglang{R} package \pkg{iglm} focuses on regression models based on joint probability models of $(\bX, \bY, \bZ)$,
where $\bX$,
$\bY$,
and $\bZ$ are random.
\proglang{R} package \pkg{iglm} allows users to specify conditional models of $(\bY, \bZ) \mid \bX$ and $(\bX, \bY) \mid \bZ$,
in addition to conditional  models of $\bY \mid \bX$ (when $\bX$ is fixed and $\bZ$ is omitted) and $\bZ \mid \bX$ (when $\bX$ is fixed and $\bY$ is omitted).
Conditional models can be specified by fixing $\bX$ or $\bZ$ using the following methods provided by \code{iglm.data}:
\begin{itemize}
\item \code{set_fix_x(TRUE)} fixes the predictors $X_i$ of all units $i$ at their observed values $x_i$, 
corresponding to a fixed covariate design and a conditional model of $(\bY, \bZ) \mid \bX$.
\item \code{set_fix_z(TRUE)} fixes all connections $Z_{i,j}$ among all pairs of units $i$ and $j$ at their observed values $z_{i,j}$, 
corresponding to a fixed network design and a conditional model of $(\bX, \bY) \mid \bZ$. 
\item \code{set_fix_z_alocal(TRUE)} fixes connections $Z_{i,j}$ among all pairs of units $i$ and $j$ without overlapping neighborhoods $\mN_i$ and $\mN_j$ at their observed values $z_{i,j}$,
but does not fix connections among units $i$ and $j$ with overlapping neighborhoods.
\end{itemize}
In line with regression,
users are free to choose a fixed design (by fixing both $\bX$ and $\bZ$) or a random design (by fixing neither $\bX$ nor $\bZ$) or a combination of a fixed design (e.g., fixing $\bX$) and a random design (e.g., not fixing $\bZ$).
Fixing $\bX$ or $\bZ$ is more restrictive than treating $\bX$ and $\bZ$ as random,
for two reasons.
First,
fixing $\bX$ and $\bZ$ at their observed values $\bx$ and $\bz$ limits conclusions to $\bx$ and $\bz$ and does not allow one to generalize conclusions to the superpopulation of all possible values of $\bX$ and $\bZ$.
In other words,
random designs afford greater generalizability than fixed designs,
at the cost of additional assumptions about the distribution of $\bX$ and $\bZ$.
Second,
fixing $\bX$ and $\bZ$ does not provide insight into the mechanism that generates predictors $\bX$ and connections $\bZ$.

In the running example,
the predictors $X_i$ are indicators of party affiliation,
which we consider to be fixed.
We fix the predictors $X_i$ at their observed values $x_i$ as follows:
\begin{verbatim}
R> data.object$set_fix_x(TRUE)
\end{verbatim}

\s 

{\bf Right-hand side of \code{formula}: terms}

\s

A comprehensive list of model terms can be found in the vignette of \proglang{R} package \pkg{iglm}:
\begin{verbatim}
R> vignette("model-terms", package = "iglm")
\end{verbatim}
These model terms include attribute-attribute,
attribute-connection,
and connection-connection terms.
Users can add custom-built model terms,
as described in Section \ref{sec:extensions}.

In the running example,
we specify the model as follows:
\label{formula.iglm}
\begin{verbatim}
R> formula.iglm <- data.object ~
+                  attribute_y + 
+                  attribute_xy + 
+                  cov_y(data = gender_attribute) +
+                  cov_y(data = white_attribute) +
+                  edges(mode = "alocal") +
+                  cov_z(data = match_gender, mode = "local") +
+                  cov_z(data = match_race, mode = "local") +
+                  cov_z(data = match_state, mode = "local") +
+                  degrees + 
+                  mutual(mode = "local") +
+                  transitive +
+                  spillover_xy(mode = "local") +
+                  spillover_yy(mode = "local") 
\end{verbatim}
We then create an instance of class \code{iglm} using 
\begin{verbatim}
R> model.iglm <- iglm(formula = formula.iglm)
\end{verbatim}
\begin{table}[t!]
    \centering
    \caption{\label{tab:suff_stats}
    Hate speech on X: 
    Model terms capturing dependence among attributes $(X_i, Y_i)$ and connections $Z_{i,j}$ in the running example. 
    Since $X_i \in \{0, 1\}$ and $Y_i \in \{0, 1\}$,
    we can write $x_i = x^\star_i$ and $y_i = y^\star_i$ for all $i \in \mP$.
    The function $d_{i,j}(\bz)$ is an indicator function,
    which is $1$ if there exists a directed two-path from unit $i$ to unit $j$ via an intermediate unit $k \in \mN_i\, \cap\, \mN_j$ and is $0$ otherwise.
    }
    \s
    \begin{tabular}{lllll}
        \toprule
Term & Statistic & Weight & Interpretation\\
\midrule 
\multicolumn{1}{l}{1. Attributes} & $g_i(\cdot)$,\, $i \in \mP$ & $\nat_g$ \\
\midrule
\code{attribute_y} & $y_i$ & $\theta_{\mY}$ & Intercept for $Y_i$\s  
\\
\code{attribute_xy} & $x_i\, y_i$ & $\theta_{\mX,\mY}$ & Predictor $X_i$ for $Y_i$\s
\\
\code{cov_y(v)} & $v_i\, y_i$ & $\theta_{\mV,\mY}$ & Predictor $v_i$ for $Y_i$
\\
\midrule
\multicolumn{1}{l}{2. Connections} & $h_{i,j}(\cdot)$,\, $(i, j) \in \mD$ & $\nat_h$\\
\midrule
\code{edges(mode = "alocal")}  & $(1-c_{i,j})\, z_{i,j}$ & $\theta_{\mZ}$ & Intercept for $Z_{i,j}$\s
\\
\code{cov_z(w, mode = "local")}  & $c_{i,j}\, w_{i,j}\, z_{i,j}$ & $\theta_{\mathscr{W},\mZ}$ & Predictor $w_{i,j}$ for $Z_{i,j}$\s
\\
\code{degrees} & $\sum_{j=1}^N z_{i,j}$ & $\theta_{\mathscr{O},i}$ & Outdegree of unit $i$\s
\\
& $\sum_{j=1}^N z_{j,i}$ & $\theta_{\mathscr{I},i}$ & Indegree of unit $i$\s
\\
\code{mutual(mode = "local")}  & $ \dfrac{c_{i,j}\, z_{i,j}\, z_{j,i}}{2}$ & $\theta_{\mZ,\mathscr{M}}$ & Mutual connection\s
\\
\code{transitive} & $ c_{i,j}\, d_{i,j}(\bz)\, z_{i,j}$ & $\theta_{\mZ,\mathscr{T}}$ & Transitive connection
\\
\midrule
\multicolumn{1}{l}{3. Attributes and Connections} & $h_{i,j}(\cdot)$,\, $(i, j) \in \mD$ & $\nat_h$
\\
\midrule
\code{spillover_xy} & $c_{i,j}\, x_i\, y_j\, z_{i,j}$ & $\theta_{\mX,\mY,\mZ}$ & Treatment spillover\s 
\\
& & & (spillover)\s 
\\
\code{spillover_yy} & $c_{i,j}\, y_i\, y_j\, z_{i,j}$ & $\theta_{\mY,\mY,\mZ}$ & Outcome spillover\s 
\\
& & & (contagion)\\
\midrule
\end{tabular}
\\
\end{table}
The definitions of model terms in the running example are presented in Table \ref{tab:suff_stats}.
Since maximum likelihood and pseudo-likelihood estimators of degree weights corresponding to isolated units without connections do not exist \citep[e.g.,][]{Ha03p,fienberg-2008},
we exclude isolated units:
\begin{verbatim}
R> data.object$delete_isolates()
\end{verbatim}

{\bf Remark: Model Terms}
The model terms used in the running example---including degree, 
transitive edge,
and regression terms capturing treatment and outcome spillover---are examples of simple and interpretable model terms.
Degree terms date back to \citet{HpLs81} and have advantages compared to alternative degree terms,
such as geometrically weighted degree terms \citep{SnPaRoHa04,HuHa04}:
\bi 
\item[1.] The geometric decay of geometrically weighted degree terms is restrictive and the decay parameter controlling the rate of decay is hard to estimate and is often fixed by users \citep[as explained in Section 1.2 of][]{StScBoMo18},
making geometrically weighted degree terms prone to misspecification.
\item[2.] 
The geometric decay of degree weights and the risk of misspecifying the decay parameter can result in inferior goodness-of-fit of models with geometrically weighted degree terms compared to models with degree terms:
e.g.,
in the running example models with degree terms turn out to be superior to models with geometrically weighted degree terms in terms of goodness-of-fit (not shown).
\item[3.] The interpretation of geometrically weighted degree terms is more complicated than the interpretation of degree terms \citep{Hu07,StScBoMo18}.
\ei 
By contrast,
the degree terms implemented in \proglang{R} package \pkg{iglm} do not impose constraints on the degree weights and help capture unobserved heterogeneity in the propensities of units to form connections.
The Minorization-Maximization (MM) algorithm described in Section \ref{sec:estimation} helps estimate degree weights without constraints.
Transitive edge terms are well-behaved alternatives to triangle terms \citep{KrKo14}.
The spillover terms capture treatment and outcome spillover.
Such terms are used in the related literature on causal inference under interference \citep[e.g.,][]{HuHa08},
peer influence \citep[e.g.,][]{manski2013identification},
and social influence \citep[e.g.,][]{KoDa22}.
In some branches of the mentioned literature,
treatment and outcome spillover are referred to as spillover and contagion,
respectively.

\s 

\hide{
{\bf Remark: Conditional and Marginal Models.}
\proglang{R} package \pkg{iglm} focuses on regression models based on joint probability models for $(\bX, \bY, \bZ)$,
where $\bX$,
$\bY$,
and $\bZ$ are all random.
\proglang{R} package \pkg{iglm} does allow users to fix $\bX$ or $\bZ$,
so that conditional models of $(\bY, \bZ) \mid \bX$ and $(\bX, \bY) \mid \bZ$ can be specified,
in addition to conditional  models of $\bY \mid \bX$ (when $\bX$ is fixed and $\bZ$ is omitted) and $\bZ \mid \bX$ (when $\bX$ is fixed and $\bY$ is omitted).
Marginal models of $\bY$ or $\bZ$ can be specified by omitting the other variables.
}

\subsection{Estimation} 
\label{sec:estimation}

To estimate the vector of weights $\nat \in \Nat$,
we maximize pseudo-likelihood functions using Minorization-Maximization and Quasi-Newton algorithms;
note that likelihood inference based on Markov chain Monte Carlo approximations of the likelihood function is feasible when $N$ is small \citep{HuHa04},
but is time-consuming when $N$ is large.

Assuming that attributes $(X_i, Y_i)$ and connections $Z_{i,j}$ are random while neighborhoods are fixed,
the pseudo-loglikelihood function based on an observation $(\bx, \by, \bz)$ of $(\bX, \bY, \bZ)$ is
\beno 
\ell(\nat; \bx, \by, \bz)
&\coloneqq& 
\underbrace{\dsum_{i \in \mathscr{P}} \ell_{\mX,i}(\nat; \bx, \by, \bz)}
&+& \underbrace{\dsum_{i \in \mathscr{P}} \ell_{\mY,i}(\nat; \bx, \by, \bz)}
&+& \underbrace{\dsum_{(i,j) \in \mathscr{D}} \ell_{\mZ,i,j}(\nat; \bx, \by, \bz).}
\\
&& \mbox{\text{~~~~\em predictors}}
&& \mbox{\text{~~~~\em outcomes}}
&& \mbox{\text{~~~~~~\em connections}}
\ee
The components $\ell_{\mX,i}(\cdot)$,
$\ell_{\mY,i}(\cdot)$,
and $\ell_{\mZ,i,j}(\cdot)$ of $\ell(\cdot)$ are the contributions of predictors,
outcomes,
and connections to the pseudo-loglikelihood function $\ell(\cdot)$:
\beno
\ell_{\mX,i}(\nat; \bx, \by, \bz)
&\coloneqq& \log f_{\nat}(x_i \mid \bx_{-i}, \by, \bz)\s
\\
\ell_{\mY,i}(\nat; \bx, \by, \bz)
&\coloneqq& \log f_{\nat}(y_i \mid \bx, \by_{-i}, \bz)\s
\\
\ell_{\mZ,i,j}(\nat; \bx, \by, \bz)
&\coloneqq& \log f_{\nat}(z_{i,j} \mid \bx, \by, \bz_{-(i,j)}).
\ee
If predictors $X_i$ or connections $Z_{i,j}$ are fixed,
we drop the corresponding components of $\ell(\cdot)$.
The pseudo-loglikelihood function $\ell(\cdot)$ is based on the conditional densities of attributes $(X_i, Y_i)$ and connections $Z_{i,j}$.
The fact that the joint probability density function of $(\bX, \bY, \bZ)$ is an exponential-family density and the functions $g_i(\cdot)$ and $h_{i,j}(\cdot)$ are coordinatewise affine functions of attributes $(X_i, Y_i)$ implies that the conditional probability density functions of $X_i$,
$Y_i$,
and $Z_{i,j}$ are exponential-family densities with normalizing constants in closed form (e.g., Bernoulli,
Poisson,
or Normal).
As a result,
$\ell(\cdot)$ is a sum of concave exponential-family loglikelihood functions,
so maximizing $\ell(\cdot)$ is a concave maximization program.

\begin{figure}
    \centering
   {\includegraphics[width=.55\textwidth,trim={9cm 10cm 0cm 0cm}, clip]{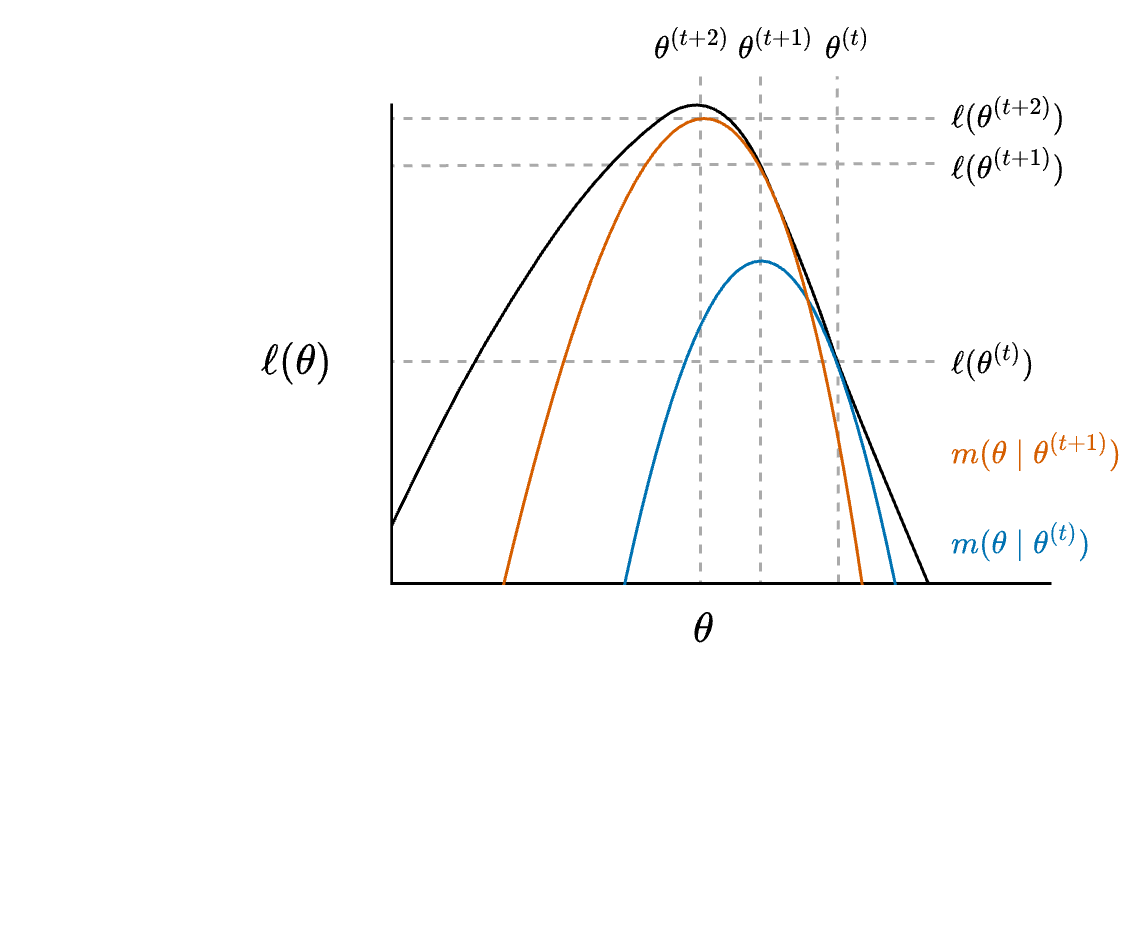}}
    \caption{A generic MM algorithm maximizing an objective function $\ell: \mR \mapsto \mR$ by maximizing a sequence of minorizing functions \textcolor[HTML]{0072B2}{$m(\nat \mid \nat^{(t)})$} and \textcolor[HTML]{D55E00}{$m(\nat \mid \nat^{(t+1)})$}.
    The dashed vertical lines represent the starting point $\nat^{(t)}$,
    the maximizer $\nat^{(t+1)}$ of \textcolor[HTML]{0072B2}{$m(\cdot \mid \nat^{(t)})$} given $\nat^{(t)}$,
    and the maximizer $\nat^{(t+2)}$ of \textcolor[HTML]{D55E00}{$m(\cdot \mid \nat^{(t+1)})$} given $\nat^{(t+1)}$.
    }
    \label{fig:placeholder}
\end{figure}

To estimate $\nat$ when the model includes the term \code{degrees},
one can partition the vector of all weights $\nat$ into the high-dimensional subvector $\nat_1$ of $2\, N$ in- and outdegree weights and the low-dimensional subvector $\nat_2$ consisting of $q$ other weights.
Iteration $t + 1$ then consists of two steps,
given interim estimates $\nat_1^{(t)}$ and $\nat_2^{(t)}$ of $\nat_1$ and $\nat_2$,
respectively:
\bi
\item[] {\bf Step 1:}
Increase $\ell(\nat_1^{(t)}, \nat_2^{(t)}; \bx, \by, \bz)$ as a function of $\nat_1^{(t)}$ for fixed $\nat_2^{(t)}$ using a Minorization-Maximization (MM) algorithm,
which can be accelerated by a Quasi-Newton algorithm,
and denote the resulting update by $\nat_1^{(t+1)}$.
\item[] {\bf Step 2:}
Increase $\ell(\nat_1^{(t+1)}, \nat_2^{(t)}; \bx, \by, \bz)$ as a function of $\nat_2^{(t)}$ for fixed $\nat_1^{(t+1)}$ using a Newton-Raphson algorithm,
and denote the resulting update by $\nat_2^{(t+1)}$.
\ei
The MM algorithm in Step 1 maximizes $\ell(\cdot)$ by maximizing a well-chosen sequence of minorizing functions.
A graphical representation of MM algorithms can be found in Figure \ref{fig:placeholder}.
The specific MM algorithm is described in \citet{FrScBhHu24}.
Taken together,
Steps 1 and 2 increase the concave pseudo-loglikelihood function $\ell(\cdot)$.
The computing time of the estimation algorithm is of order $O(N^2\, q^2 + q^3)$ when $O(N)$ degree terms and $q$ other model terms are used,
where $O(N^2)$ is the cost of Step 1 and $O(N^2\, q^2 + q^3)$ is the cost of Step 2.
The resulting pseudo-likelihood estimators are supported by provable theoretical guarantees---including consistency results and rates of convergence---and simulation results \citep{FrScBhHu24}.

\paragraph*{Example}

The estimation algorithm can be specified via the object \code{iglm.control}.
In the running example,
we use the following options to specify the estimation algorithm:
\begin{verbatim}
R> control.obj <- control.iglm(max_it = 300)
\end{verbatim}
The option \code{max_it = 300} limits the number of iterations of the estimation algorithm to 300.
Additional options allow users to specify which information the estimation algorithm returns.
A comprehensive list of available options can be found in the help function of \proglang{R} package \pkg{iglm}:
see \code{?control.iglm}.

\begin{figure}[t!]
\centering
\includegraphics[width=0.45\textwidth]{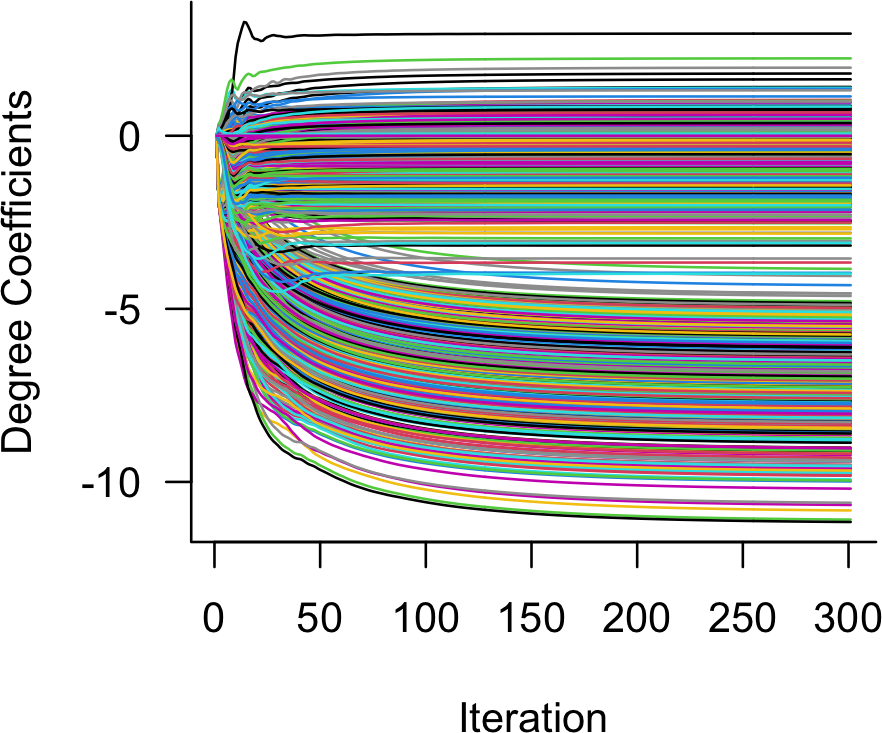}
\includegraphics[width=0.45\textwidth]{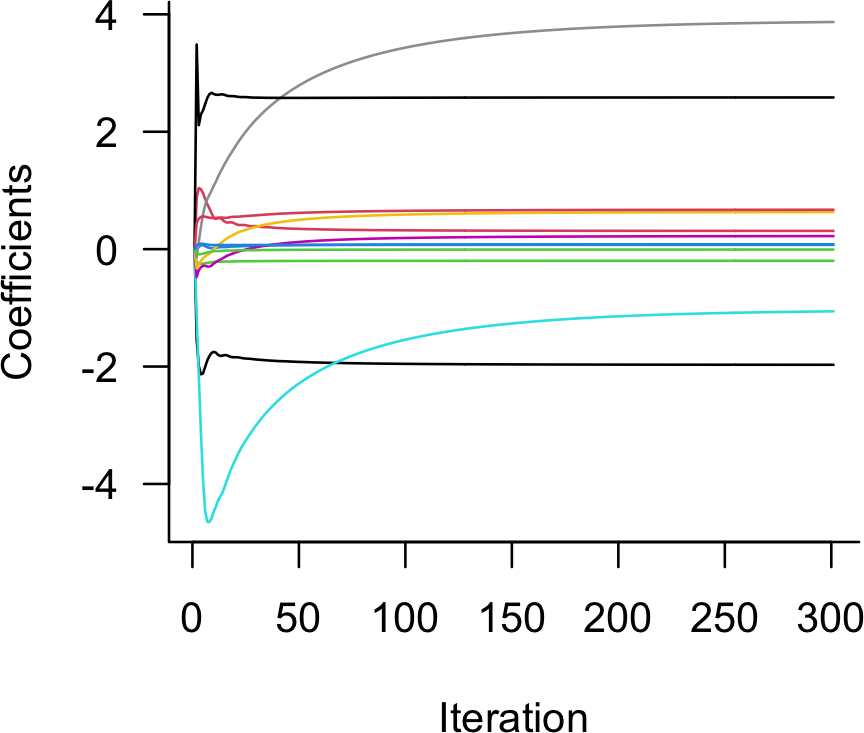}
\caption{Hate speech on X: 
Trace plots of estimates,
including estimates of degree weights (left) and all other weights (right).}
\label{fig:trace}
\end{figure}

We use all options specified in \code{iglm.control} for the \code{model.iglm} object using
\begin{verbatim}
R> model.iglm$set_control(control.obj)
\end{verbatim}
We then call the function \code{estimate()} to estimate the model: 
\begin{verbatim}
R> model.iglm$estimate()
\end{verbatim}
Output generated by operations on an \code{iglm} object is stored in \code{results},
which is a \code{R6} class containing methods to display diagnostics, 
including trace plots of estimates:
\begin{verbatim}
R> model.iglm$results$plot(trace = TRUE)
\end{verbatim}
Trace plots of estimates help detect non-convergence of the estimation algorithm.
In the running example,
the trace plots in Figure \ref{fig:trace} do not reveal signs of non-convergence.

\subsection{Uncertainty}
\label{sec:uq}

We approximate the covariance matrix of maximum pseudo-likelihood estimator $\widehat\nat$ as follows.

First,
the mean-value theorem for vector-valued functions \citep[][Equations (2) and (3), pp.~68--69]{OrRh00} implies that there exist real numbers $t_1, \ldots, t_p \in (0,\, 1)$ such that the exact covariance matrix of $\widehat\nat$ based on fixed and finite $N$ is
\beno
\mbV_{N,\truth}(\widehat\nat)
&=& \mbV_{N,\truth}\left[-\bm{H}(\widehat\nat,\, \truth; \bX, \bY,  \bZ)^{-1}\; \nabla_{\nat}\; \ell(\nat; \bX, \bY,  \bZ)\Big|_{\nat=\truth}\right],
\ee
where $\truth \in \mR^p$ is the data-generating vector of weights.
The matrix $\bm{H}(\cdot) \in \mR^{p \times p}$ is
\beno
\bm{H}(\widehat\nat,\, \truth; \bX, \bY,  \bZ)
&\coloneqq&
\left(
\begin{array}{cccccc}
u_1^\prime(\truth + t_1\, (\widehat\nat - \truth);\, \bX, \bY,  \bZ)
\\
\vdots
\\
u_p^\prime(\truth + t_p\, (\widehat\nat - \truth);\, \bX, \bY,  \bZ)
\end{array}
\right),
\ee
where $u_k(\cdot)$ is the $k$th coordinate of $\nabla_{\nat}\,\, \ell(\cdot)$ and $u_k^\prime(\cdot)$ is the row vector of partial derivatives of $u_k(\cdot)$ with respect to $\nat$ ($k = 1, \ldots, p$).

Second,
we follow convention and replace $\truth$ by $\widehat\nat$,
motivated by the observation that\break 
$\max_{1 \leq i \leq p} |\widehat\theta_i - \theta_i^\star|$ is small with high probability for all large enough $N$.
For example,
Corollary 1 of \citet[]{FrScBhHu24} shows that there exist constants $C > 0$ and $N_0 > 0$ such that
\beno
\mbP\left(\max\limits_{1 \leq i \leq p}\, |\widehat\theta_i - \theta_i^\star|
\;<\; C\, \sqrt{\dfrac{\log N}{N}}\right)
&\geq& 1 - \dfrac{6}{N^2}
&\mbox{for all}& 
N > N_0.
\ee
As a result,
$\widehat\nat$ is expected to be a reasonable approximation of $\truth$ for all $N > N_0$.
Upon replacing $\truth$ by $\widehat\nat$,
we obtain the approximate covariance matrix
\beno
\label{approximate.covariance}
\mbV_{N,\widehat\nat}(\widehat\nat)
&=& \mbV_{N,\widehat\nat}\left[-\nabla_{\nat}^2\; \ell(\nat; \bX, \bY,  \bZ)\Big|_{\nat=\widehat\nat}^{-1}\;\, \nabla_{\nat}\; \ell(\nat; \bX, \bY,  \bZ)\Big|_{\nat=\widehat\nat}\right].
\ee

We use Markov chain Monte Carlo simulations of attributes $(X_i, Y_i)$ and connections $Z_{i,j}$ to approximate $\mbV_{N,\widehat\nat}(\widehat\nat)$,
as described in Section \ref{sec:simulation}.
The standard errors of $\widehat\nat$ are the square roots of the diagonal elements of $\mbV_{N,\widehat\nat}(\widehat\nat)$.
In the running example, 
the standard errors of $\widehat\nat$ are shown in the second column of the summary reported in Section \ref{sec:interpretation}.

\subsection{Simulation}
\label{sec:simulation}

To quantify the uncertainty about maximum pseudo-likelihood estimator $\widehat\nat$ and assess models, 
we simulate attributes $(X_i, Y_i)$ and connections $Z_{i,j}$ using Markov chain Monte Carlo methods.

If both $(X_i, Y_i)$ and $Z_{i,j}$ are random,
we can simulate $(X_i, Y_i)$ and $Z_{i,j}$ using Gibbs sampling:
\begin{enumerate}
\item Sample $X_i \mid (\bX_{-i}, \bY, \bZ) = (\bx_{-i}, \by, \bz)$ for all units $i \in \mathscr{P}$.
\item Sample $Y_i \mid (\bX, \bY_{-i}, \bZ) = (\bx, \by_{-i}, \bz)$ for all units $i \in \mathscr{P}$.
\item Sample $Z_{i,j} \mid (\bX, \bY, \bZ_{-(i,j)}) = (\bx, \by, \bz_{-(i,j)})$ for all pairs of units $(i,j )\in \mathscr{D}$.
\end{enumerate}
An alternative to simulating connections $Z_{i,j}$ by a Gibbs sampler is a Metropolis-Hastings algorithm with TNT (Tie-No-Tie) proposals,
which selects an unconnected or connected pair of units with equal probability and proposes toggling it.
Metropolis-Hastings algorithms with TNT proposals are popular in \proglang{R} packages for dependent connections $Z_{i,j}$,
including \code{ergm} \citep{ergm2023},
and can improve the mixing of Markov chain Monte Carlo algorithms for simulating sparse networks.

In the running example,
outcomes $Y_i$ and connections $Z_{i,j}$ are random while predictors $X_i$ are fixed,
because the predictors $X_i$ are indicators of party affiliation.
We specify Markov chain Monte Carlo algorithms for sampling $Y_i$ and $Z_{i,j}$ as follows: 
\begin{verbatim}
R> sampler.y.obj <- sampler.net.attr(n_proposals = data.object$n_units * 10)
R> M <- nrow(data.object$overlap) * 10
R> sampler.z.obj <- sampler.net.attr(n_proposals = M, tnt = TRUE)
\end{verbatim}
The object \code{sampler.y.obj} specifies a Gibbs sampler for simulating $Y_i$.
The object\break 
\code{sampler.z.obj} specifies a Metropolis-Hastings algorithm with TNT proposals for simulating $Z_{i,j}$.
These Markov chain Monte Carlo algorithms can be combined as follows:
\begin{verbatim}
R> sampler.obj <- sampler.iglm(n_burn_in = 1, 
+                              n_simulation = 1, 
+                              seed = 123,
+                              sampler_y = sampler.y.obj, 
+                              sampler_z = sampler.z.obj)
\end{verbatim}
The options \code{n_burn_in} and \code{n_simulation} specify the numbers of burn-in and post-burn-in iterations.
Using \code{sampler.obj},
we can then simulate $Y_i$ and $Z_{i,j}$ as follows:
\begin{verbatim}
R> model.iglm$set_sampler(sampler.obj)
R> model.iglm$simulate()
\end{verbatim}
The resulting samples are saved in \code{model.iglm$results$samples} as an \code{iglm.data.list}. 

Sometimes, 
users may wish to simulate from a specified model.
To do so,
users need to specify the model in Section \ref{sec:specification} along with the values of all weights.
We can specify the values of all weights by specifying the values \code{coef_degree} of the degree weights (when the term \code{degrees} is used) and the values \code{coef} of all other weights:
\begin{verbatim}
R> model.sim <- iglm(formula = formula.iglm, 
+                    sampler = sampler.obj,
+                    coef = model.iglm$coef, 
+                    coef_degree = model.iglm$coef_degrees)
R> model.sim$simulate()
\end{verbatim}
The function \code{simulate()} simulates attributes $(X_i, Y_i)$ and connections $Z_{i,j}$.
In the above call, 
we do not specify the number of Markov chain Monte Carlo draws but use the default of 100 draws,
starting from the observed data stored in \code{iglm.data}.

\subsection{Assessment}
\label{sec:assessment}

To assess models,
\proglang{R} package \pkg{iglm} compares simulated and observed data.
To do so,
we call the function \code{assess()}. 
Its arguments specify the statistics to be computed.

\begin{figure}[t!]
  \centering
  \includegraphics[width=0.45\textwidth, clip]{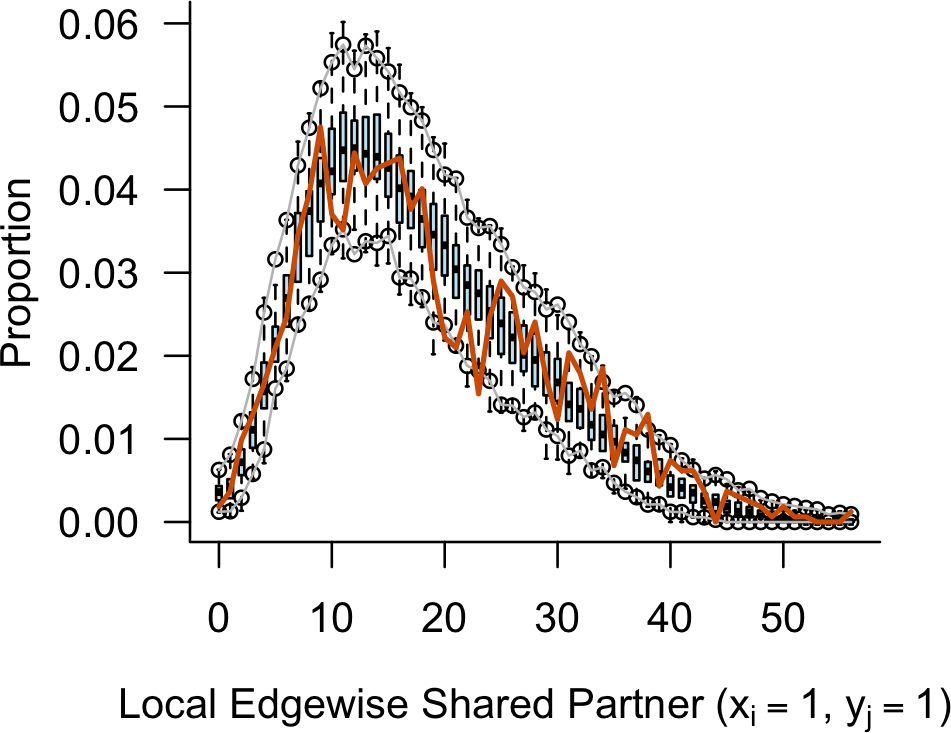}
  \includegraphics[width=0.45\textwidth, clip]{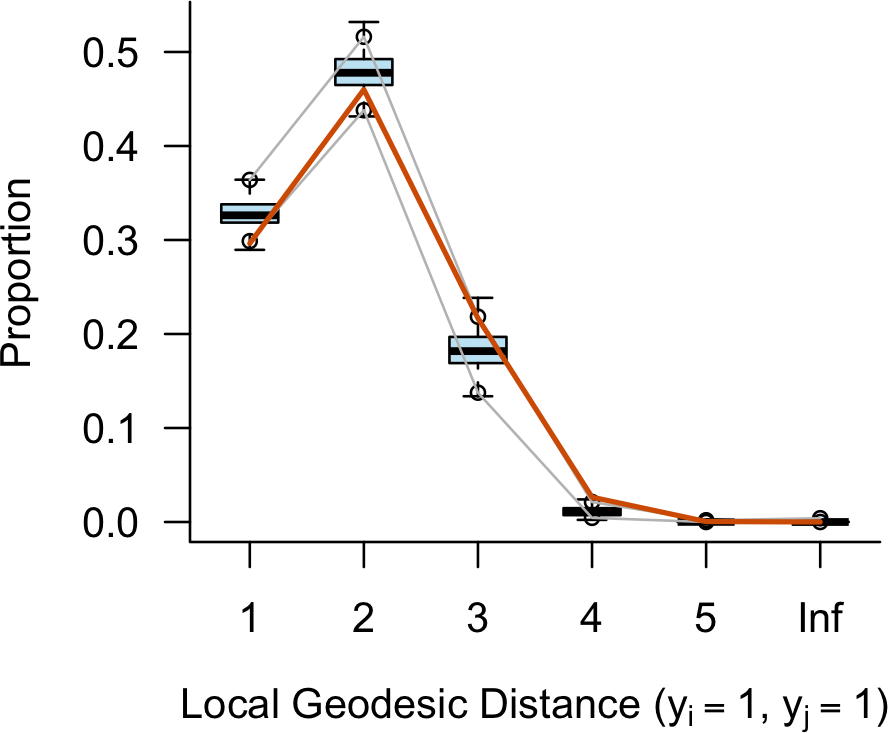}
  \s\s
  
  \includegraphics[width=0.45\textwidth, clip]{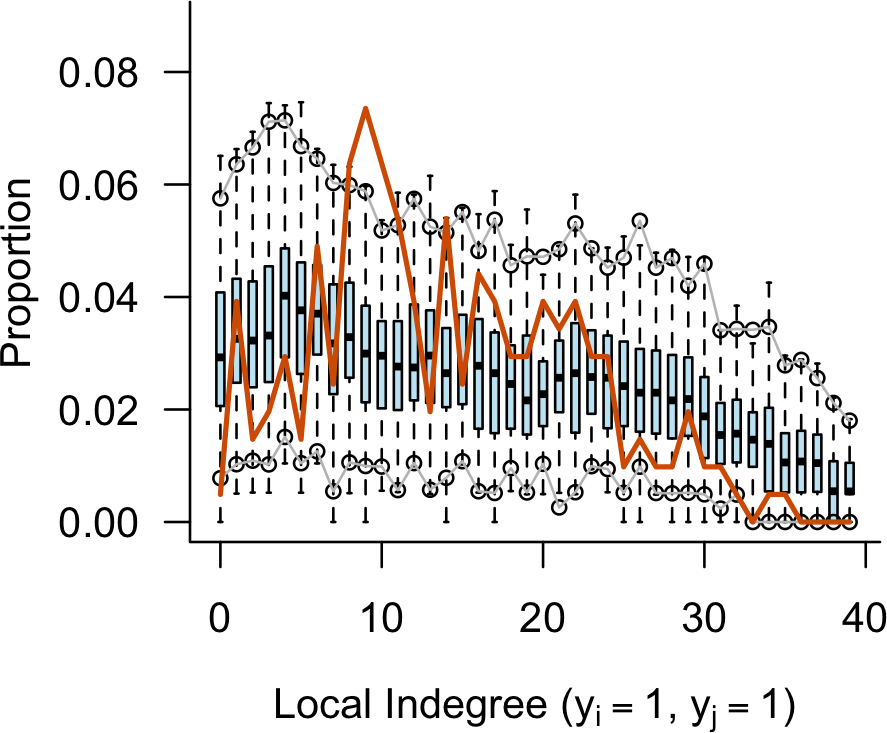}
  \includegraphics[width=0.45\textwidth, clip]{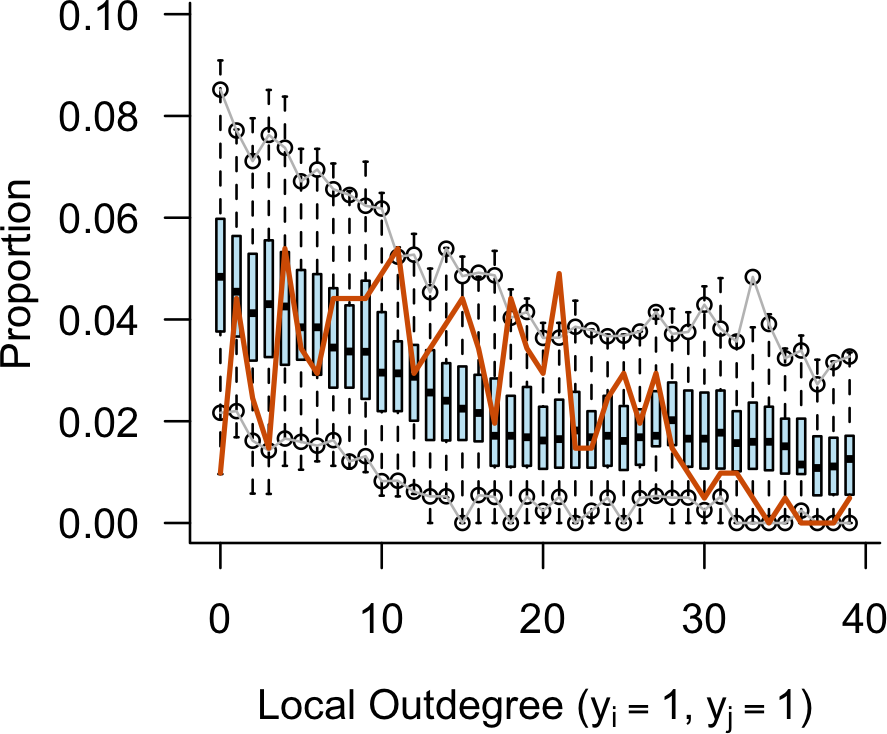}
   \caption{Hate speech on X: 
   Model assessment in terms of the numbers of local edgewise shared partners among pairs of units $\{i, j\}$ with $c_{i,j}\, x_i\, y_j = 1$ or $c_{i,j}\, x_j\, y_i = 1$,
   local geodesic distances among pairs of units $\{i, j\}$ with $c_{i,j}\, y_i\, y_j = 1$,
   and local spillover in- and outdegrees in the subnetwork consisting of pairs of units $\{i, j\}$ with $c_{i,j}\, y_i\, y_j = 1$.
   }
  \label{fig:model_assessment}
\end{figure}

\paragraph*{Example}
In the running example,
we assess the model as follows:
\begin{verbatim}
R> model.iglm$assess(formula = ~ esp_dist(x_i = 1, y_j = 1, mode = "local") + 
+                    geo_dist(y_i = 1, y_j = 1, mode = "local") +
+                    deg_dist(y_i = 1, y_j = 1, mode = "local"))
\end{verbatim}
The function \code{assess()} searches in object \code{model.iglm} for existing simulations.
If \code{model.iglm} contains existing simulations,
it will use them.
Otherwise,
it will simulate data.
In either case,
it will plot the results,
as demonstrated in Figure \ref{fig:model_assessment}. 
Figure \ref{fig:model_assessment} suggests that model-based predictions match the empirical distributions of the numbers of local edgewise shared partners, 
geodesic distances, 
and spillover in- and outdegrees. 

\subsection{Interpretation } 
\label{sec:interpretation}

Models can be interpreted by inspecting the conditional distributions of attributes $(X_i, Y_i)$ and connections $Z_{i,j}$.
The conditional distributions of $(X_i, Y_i)$ and $Z_{i,j}$ can be represented by GLMs as long as the vector-valued functions $g_{i}(\cdot)$ and $h_{i,j}(\cdot)$ are coordinatewise affine functions of $(X_i, Y_i)$ \citep[Proposition 1,][]{FrScBhHu24}.
The GLM representations of conditionals facilitate interpretation,
because GLMs are widely used in data science.

To demonstrate,
let $g_{i}(\cdot)$ and $h_{i,j}(\cdot)$ be coordinatewise affine functions of $(X_i, Y_i)$,
in the sense that there exist vector-valued functions 
$g_{\mX,i,k}(\cdot)$,
$g_{\mY,i,k}(\cdot)$,
$h_{\mX,i,j,k}(\cdot)$, and 
$h_{\mY,i,j,k}(\cdot)$ ($k = 0, 1, 2, 3$) such that
\begin{equation}
\label{eq:lp}
\begin{array}{clllll}
g_i(x_i^\star, y_i^\star) 
&=& g_{\mX,i,0}(y_i^\star) + g_{\mX,i,1}(y_i^\star)\, x_i^\star\s
\\
&=& g_{\mY,i,0}(x_i^\star) + g_{\mY,i,1}(x_i^\star)\, y_i^\star\s
\\
h_{i,j}(x_i^\star, x_j^\star, y_i^\star, y^\star_j, \bz) 
&=& h_{\mX,i,j,0}(y_i^\star,y^\star_j,\bz) + h_{\mX,i,j,1}(y_i^\star,y^\star_j,\bz)\, x_i^\star + h_{\mX,i,j,2}(y_i^\star,y^\star_j,\bz)\, x_j^\star\s
\\
&+& h_{\mX,i,j,3}(y_i^\star,y^\star_j,\bz)\, x_i^\star\, x_j^\star\s
\\
&=& h_{\mY,i,j,0}(x_i^\star,x_j^\star,\bz) + h_{\mY,i,j,1}(x_i^\star,x_j^\star,\bz)\, y_i^\star + h_{\mY,i,j,2}(x_i^\star,x_j^\star,\bz)\, y^\star_j\s
\\
&+& h_{\mY,i,j,3}(x_i^\star,x_j^\star,\bz)\, y_i^\star\, y^\star_j.
\end{array}
\end{equation}
We assume that $h_{j,i}(\cdot) \coloneqq 0$ when connections are undirected,
because the joint probability density function of $(\bX, \bY, \bZ)$ stated in Section \ref{sec:model} depends on $h_{i,j}(\cdot)$ but does not depend on $h_{j,i}(\cdot)$ ($i < j$).
Extending Proposition 1 of \citet{FrScBhHu24} from $Y_i$ to $X_i$ and $Y_i$ shows that the conditional distributions of $X_i$ and $Y_i$ can be represented by GLMs with linear predictors
\be
\label{eta.eq}
\eta_{\mX,i}
&\coloneqq& \nat^\top \left(g_{\mX,i,1}(y_i^\star),\; h_{\mX,i}(\bm{x}_{-i}^\star, \bm{y}^\star, \bz)\right)
&=& \nat^\top\, \Delta_{\mX,i}\s\s
\\ 
\eta_{\mY,i}
&\coloneqq& \nat^\top \left(g_{\mY,i,1}(x_i^\star),\; h_{\mY,i}(\bm{x}^\star, \bm{y}_{-i}^\star, \bz)\right)
&=& \nat^\top\, \Delta_{\mY,i}.
\ee
The functions $h_{\mX,i}(\cdot)$ and $h_{\mY,i}(\cdot)$ are defined by
\beno
h_{\mX,i}(\bm{x}_{-i}^\star, \bm{y}^\star, \bz)
\coloneqq \dsum_{j \in \mP \setminus \{i\}}\, \left[h_{\mX,i,j,1}(\cdot) + h_{\mX,j,i,2}(\cdot) + h_{\mX,i,j,3}(\cdot)\, x_j^\star + h_{\mX,j,i,3}(\cdot)\, x_j^\star\right]\s 
\\
h_{\mY,i}(\bm{x}^\star, \bm{y}_{-i}^\star, \bz)
\coloneqq \dsum_{j \in \mP \setminus \{i\}}\, \left[h_{\mY,i,j,1}(\cdot) + h_{\mY,j,i,2}(\cdot) + h_{\mY,i,j,3}(\cdot)\, y^\star_j + h_{\mY,j,i,3}(\cdot)\, y^\star_j \right],
\ee
where the arguments of functions on the right-hand side are suppressed for the sake of brevity.
We demonstrate how the conditional distribution of $Y_i$ can be represented by a GLM in the running example.
First,
the conditional distribution of $Y_i \in \{0, 1\}$ is Bernoulli with mean 
\beno
\mu_{\mY,i}(\eta_{\mY,i})
\;\coloneqq\; \mbE_{\eta_{\mY,i}}[Y_i \mid (\bX, \bY_{-i}, \bZ) = (\bx, \by_{-i}, \bz)]
\;=\; \dfrac{\exp(\eta_{\mY,i})}{1 + \exp(\eta_{\mY,i})}.
\ee
Second,
the linear predictor $\eta_{\mathscr{Y},i}$ can be derived by starting with the first term in the formula \code{formula.iglm} stated in Section \ref{sec:specification} and adding the other terms one by one:
\bi
\item[]1. \code{attribute_y} implies that $g_i(\cdot)$ includes coordinate $y_i$.
Since the term $y_i$ with weight $\theta_{\mY}$ is a linear function of $y_i$,
\code{attribute_y} adds $\theta_{\mY}$ to linear predictor $\eta_{\mY,i}$.
\item[]2. Adding \code{attribute_xy} adds coordinate $x_i\, y_i$ to $g_i(\cdot)$.
Since the term $x_i\, y_i$ with weight $\theta_{\mX,\mY}$ is a linear function of $y_i$,
\code{attribute_xy} adds $\theta_{\mX,\mY}\, x_i$ to linear predictor $\eta_{\mY,i}$.
\item[]3--4. Adding \code{cov_y} with predictors \code{gender_attribute} and \code{white_attribute} adds coordinates $v_{i,1}\, y_i$ and $v_{i,2}\, y_i$ to $g_i(\cdot)$.
The terms $v_{i,1}\, y_i$ and $v_{i,2}\, y_i$ with weights $\theta_{\mV,\mY,1}$ and $\theta_{\mV,\mY,2}$ are linear functions of $y_i$ and add 
$\theta_{\mV,\mY,1}\, v_{i,1} + \theta_{\mV,\mY,2}\, v_{i,2}$ to linear predictor $\eta_{\mY,i}$.
\hide{
\beno
\eta_{\mY,i} 
&=& \theta_{\mY} + \theta_{\mX,\mY}\, x_i + \theta_{\mV,\mY,1}\, v_{i,1} + \theta_{\mV,\mY,2}\, v_{i,2}.
\ee
}
\item[]5. Adding \code{spillover_xy} implies that $h_{i,j}(\cdot)$ and $h_{j,i}(\cdot)$ consist of $c_{i,j}\, x_i\, y_j\, z_{i,j}$ and $c_{j,i}\, x_j\, y_i\, z_{j,i}$,
each of them with weight $\theta_{\mX,\mY,\mZ}$.
Both terms are linear functions of $y_i$,
adding $\theta_{\mX,\mY,\mZ} \sum_{j \in \mP \setminus \{i\}} c_{i,j}\, x_j\, z_{j,i}$ to linear predictor $\eta_{\mY,i}$;
note that $c_{i,j} = c_{j,i}$.
\item[]6. Adding \code{spillover_yy} adds $c_{i,j}\, y_i\, y_j\, z_{i,j}$ and $c_{j,i}\, y_j\, y_i\, z_{j,i}$ to $h_{i,j}(\cdot)$ and $h_{j,i}(\cdot)$,
each with weight $\theta_{\mY,\mY,\mZ}$.
Both are linear functions of $y_i$,
adding $\theta_{\mY,\mY,\mZ} \sum_{j \in \mP \setminus \{i\}} c_{i,j}\, y_j\, (z_{i,j} + z_{j,i})$ to linear predictor $\eta_{\mY,i}$.
\ei
All other terms in \code{formula.iglm} are not functions of $y_i$ and hence do not affect the conditional distribution of $Y_i$.
Upon collecting terms,
the linear predictor turns out to be
\[
\begin{array}{lllllllllllll}
\eta_{\mY,i}
\;\coloneqq\; \theta_{\mY} 
\;+\; \theta_{\mX,\mY}\, x_{i} 
\;+\; \theta_{\mV,\mY,1}\, v_{i,1} 
\;+\; \theta_{\mV,\mY,2}\, v_{i,2} 
&+& \theta_{\mX,\mY,\mZ} \dsum_{j \in \mP \setminus \{i\}} c_{i,j}\, x_j\, z_{j,i}\s 
\\
&+& \theta_{\mY,\mY,\mZ} \dsum_{j \in \mP \setminus \{i\}} c_{i,j}\, y_j\, (z_{i,j} + z_{j,i}).
\end{array}
\]

The conditional distribution of $Z_{i,j}$ can be represented by a GLM with linear predictor 
\be
\label{change.z}
\eta_{\,\mathscr{Z},i,j}
&\coloneqq& \nat_h^\top \left(s(\bx, \by, z_{i,j} = 1, \bz_{-(i,j)}) - s(\bx, \by, z_{i,j} = 0, \bz_{-(i,j)})\right)
&=& \nat_h^\top\, \Delta_{\,\mathscr{Z},i,j},
\ee
where
\beno
s(\bx, \by, z_{i,j}, \bz_{-(i,j)})
&\coloneqq& \dsum_{(a,b) \in \mD}\, h_{a,b}(x_a, x_b, y_a, y_b, z_{i,j}, \bz_{-(i,j)}).
\ee

The conditional distribution of $Z_{i,j} \in \{0, 1\}$ is Bernoulli with mean
\beno
\mu_{\mZ,i,j}(\eta_{\mZ,i,j})
\;\coloneqq\; \mbE_{\eta_{\mZ,i,j}}[Z_{i,j} \mid (\bX, \bY, \bZ_{-(i,j)}) = (\bx, \by, \bz_{-(i,j)})]
\;=\; \dfrac{\exp(\eta_{\mZ,i,j})}{1 + \exp(\eta_{\mZ,i,j})},
\ee
where
\beno 
\eta_{\mZ,i,j}
&\coloneqq& \theta_{\mathscr{O},i}
+ \theta_{\mathscr{I},j} 
+ \theta_{\mZ}\, (1 - c_{i,j})\s 
\\
&+& \left[
\theta_{\mV,\mZ,1}\, \mathbb{I}(v_{i,1} = v_{j,1})
+ \theta_{\mV,\mZ,2}\, \mathbb{I}(v_{i,2} = v_{j,2})
+ \theta_{\mV,\mZ,3}\, \mathbb{I}(v_{i,3} = v_{j,3})\right]\, c_{i,j}\s 
\\
&+& \left[\theta_{\mZ,\mathscr{M}}\, 
z_{j,i}
\;+\; \theta_{\mZ,\mathscr{T}}\, \Delta_{\mZ,i,j}(\bz)
\;+\; \theta_{\mX,\mY,\mZ}\, x_i\, y_j
\;+\; \theta_{\mY,\mY,\mZ}\, y_i\, y_j \right]\, c_{i,j}.
\ee
Here,
$\Delta_{\mZ,i,j}(\bz) \coloneqq \sum_{(a,b) \in \mD}\, [d_{a,b}(z_{i,j} = 1, \bz_{-(i,j)}) - d_{a,b}(z_{i,j} = 0, \bz_{-(i,j)})]$,
where $d_{a,b}(\bz)$ is $1$ if units $a$ and $b$ are both connected to one or more units $k \in \mN_a \cap \mN_b$ and is $0$ otherwise.

\paragraph{Example}
Using the function \code{summary()},
we obtain the following maximum pseudo-likelihood estimate $\widehat\nat$ and standard errors:
\begin{verbatim}
R> model.iglm$summary()
iglm object
----------------------------------------------------------------------------
Results: 

                                           Estimate    S.E. t-value Pr(>|t|)
attribute_y                                 -1.9697  0.2575   -7.65  <0.0001
attribute_xy                                 0.3109  0.2455    1.27     0.21
cov_y(data = gender_attribute)              -0.1986  0.2626   -0.76     0.45
cov_y(data = white_attribute)                0.0825  0.2754    0.30     0.76
edges(mode = 'alocal')                      -1.0593  0.1625   -6.52  <0.0001
cov_z(data = match_gender, mode = 'local')   0.2223  0.0313    7.11  <0.0001
cov_z(data = match_race, mode = 'local')     0.6322  0.0326   19.40  <0.0001
cov_z(data = match_state, mode = 'local')    3.8695  0.1580   24.48  <0.0001
mutual(mode = 'local')                       2.5842  0.0576   44.86  <0.0001
transitive                                   0.6712  0.0772    8.69  <0.0001
spillover_xy                                -0.0069  0.0198   -0.35     0.73
spillover_yy                                 0.0732  0.0079    9.22  <0.0001

Time for estimation: 8.2 mins

Degree Parameters:
  Outdegrees:
   Min. 1st Qu.  Median    Mean 3rd Qu.    Max. 
  -11.2    -7.8    -6.9    -7.1    -6.2    -3.8 

  Indegrees:
   Min. 1st Qu.  Median    Mean 3rd Qu.    Max. 
 -3.983  -1.126  -0.571  -0.628  -0.046   2.951 
\end{verbatim}

Since there are $2\, N = 990$ in- and outdegree weights,
the function \code{summary()} reports a summary of in- and outdegree weights,
but it does not print all $990$ estimates.
Users can access the estimates of all in- and outdegree weights using \code{model.iglm$coef_degrees}.
The tail probabilities \code{Pr(>|t|)} are based on $N(0, 1)$ distributions with mean $0$ and variance $1$.
We discuss non-asymptotic and asymptotic approximations of tail probabilities in Section \ref{sec:discussion}.

The conditional distribution of outcome $Y_i \in \{0, 1\}$ is Bernoulli with mean 
\beno
\mu_{\mY,i}(\widehat\eta_{\mY,i})
\;\coloneqq\; \mbE_{\widehat\eta_{\mY,i}}[Y_i \mid (\bX, \bY_{-i}, \bZ) = (\bx, \by_{-i}, \bz)]
\;=\; \dfrac{\exp(\widehat\eta_{\mY,i})}{1 + \exp(\widehat\eta_{\mY,i})},
\ee
where
\[
\begin{array}{lllllllllllll}
\widehat\eta_{\mY,i}
&\coloneqq& -1.970
\;+.311\, x_{i} 
\;-.199\, v_{i,1} 
\;+.083\, v_{i,2} \s\\ 
&-& .007 \dsum_{j \in \mP \setminus \{i\}} c_{i,j}\, x_j\, z_{j,i}\s 
\;+ .073 \dsum_{j \in \mP \setminus \{i\}} c_{i,j}\, y_j\, (z_{i,j} + z_{j,i}).
\end{array}
\]
These results indicate that demographic background in terms of party affiliation,
gender,
and race
does not help predict hate speech.
The treatment spillover effect is not significant at conventional significance levels (e.g., $.05$),
whereas the outcome spillover effect is significant.
To interpret the outcome spillover term,
consider any legislator $i$.
Then each legislator $j$ who is close to $i$ ($c_{i,j} = 1$),
connected to $i$ (either $z_{i,j} = 1$ or $z_{j,i} = 1$),
and uses hate speech ($y_j = 1$) adds $.073$ to the log odds of the conditional probability that $i$ uses hate speech ($Y_i = 1$).

The conditional distribution of connection $Z_{i,j} \in \{0, 1\}$ is Bernoulli with mean
\beno
\mu_{\mZ,i,j}(\widehat\eta_{\mZ,i,j})
\;\coloneqq\; \mbE_{\widehat\eta_{\mZ,i,j}}[Z_{i,j} \mid (\bX, \bY, \bZ_{-(i,j)}) = (\bx, \by, \bz_{-(i,j)})]
\;=\; \dfrac{\exp(\widehat\eta_{\mZ,i,j})}{1 + \exp(\widehat\eta_{\mZ,i,j})},
\ee
where
\beno 
\widehat\eta_{\mZ,i,j}
&\coloneqq& \widehat\theta_{\mathscr{O},i}
+ \widehat\theta_{\mathscr{I},j} 
-1.059\, (1 - c_{i,j})\s 
\\
&+& \left[.222\, \mathbb{I}(v_{i,1} = v_{j,1})
+ .632\, \mathbb{I}(v_{i,2} = v_{j,2})
+ 3.870\, \mathbb{I}(v_{i,3} = v_{j,3})\right]\, c_{i,j}\s 
\\
&+& \left[2.584\, z_{j,i}
+ .671\, \Delta_{\mZ,i,j}(\bz)
- .007\, x_i\, y_j
+ .073\, y_i\, y_j \right]\, c_{i,j}.
\ee
The estimates $\widehat\theta_{\mathscr{O},i}$ and $\widehat\theta_{\mathscr{I},j}$ can be extracted from \code{model.iglm$coef_degrees}. 
The above results indicate a tendency to connect with others who are similar in terms of hate speech,
in addition to reciprocating connections (mutual connections) and turning indirect connections into direct ones (transitive connections),
which dovetails with findings in network analysis \citep[e.g.,][]{WsFk94,ergm.book}.

\subsection{Misspecification of Neighborhoods}
\label{sec:misspecification}

\begin{figure}[t!]
  \centering
  \includegraphics[width=\textwidth]{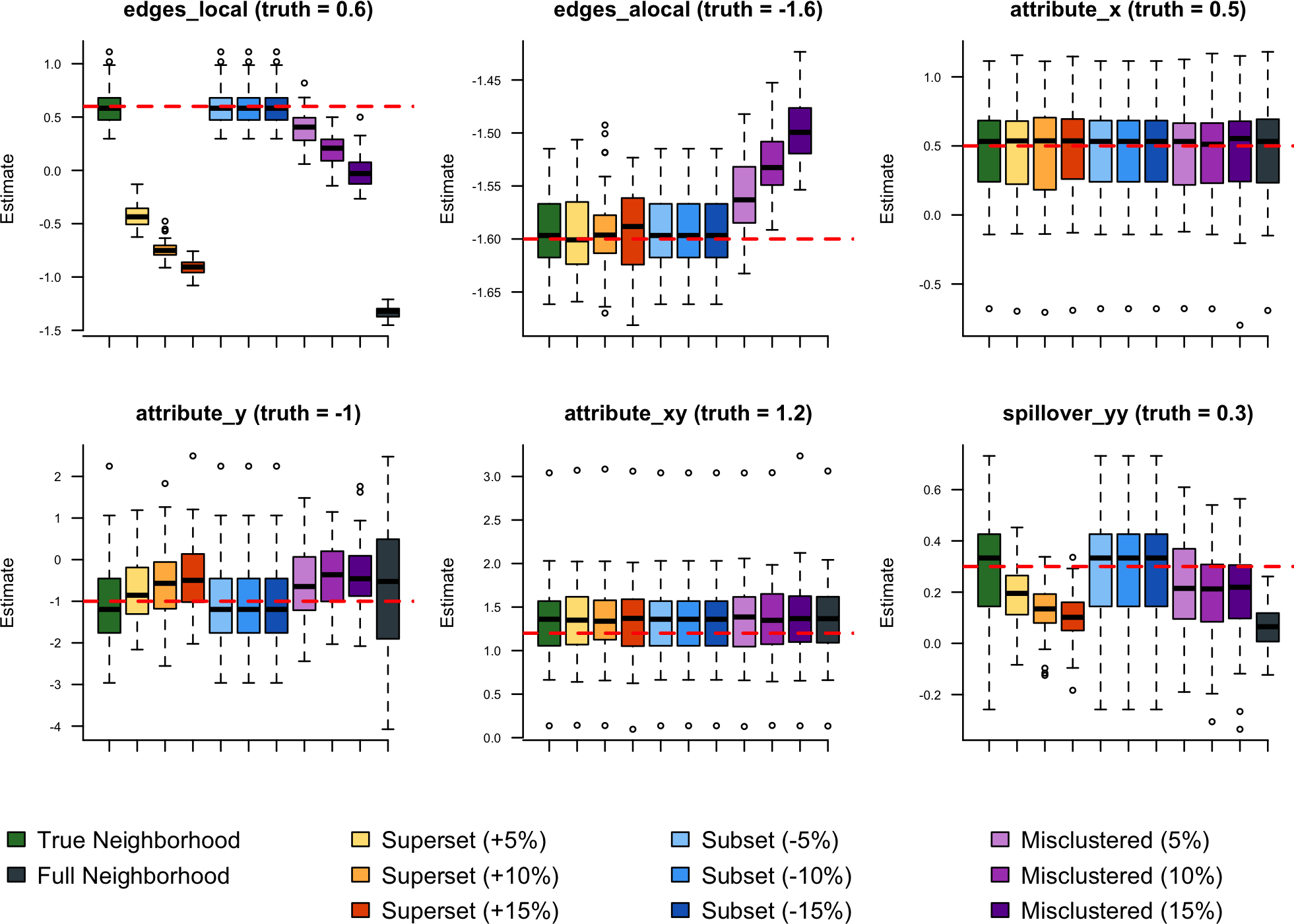}
   \caption{Simulation results when specifications $\mN_i$ of neighborhoods $\mN_i^\star$ are correct (``True Neighborhood'') or incorrect.
   Incorrect specifications include $\mN_i = \mP$ (``Full Neighborhood''),
   supersets $\mN_i \supset \mN_i^\star$ (``Superset''),
   subsets $\mN_i \subset \mN_i^\star$ (``Subset''),
   and $\mN_i \neq \mN_i^\star$ (``Misclustered'').}
  \label{fig:simul_study}
\end{figure}

We assess the robustness of maximum pseudo-likelihood estimator $\widehat\nat$ to misspecifications of neighborhoods using simulations.
We assume that the population $\mP \coloneqq \{1, \ldots, 100\}$ consists of $10$ non-overlapping subpopulations $\{1, \ldots, 10\}, \ldots, \{91, \ldots, 100\}$ of size $10$,
and the neighborhoods of units $i \in \mP$ consist of all other units in the same subpopulation.
The data-generating model is described in Appendix \ref{sec:simulation_study}.
We consider the following forms of misspecification,
where $\mN_i^\star$ denotes the true neighborhood and $\mN_i$ denotes the specified neighborhood of unit $i \in \mP$:
\bi
\item full neighborhoods $\mN_i = \mP$;
\item supersets $\mN_i \supset \mN_i^\star$,
i.e.,
the specified neighborhoods $\mN_i$ consist of all true neighbors $j \in \mN_i^\star$ and 5\%,
10\%,
and 15\% non-neighbors $j \in \mP \setminus \mN_i^\star$;
\item subsets $\mN_i \subset \mN_i^\star$,
i.e.,
the specified neighborhoods $\mN_i$ are obtained by taking the set of all true neighbors $j \in \mN_i^\star$ and removing 5\%,
10\%,
and 15\% of the true neighbors $j \in \mN_i^\star$;
\item misclustering,
i.e.,
we sample at random $5\%, 10\%,$ or $15\%$ of units and move each of them from their home subpopulation to some other subpopulation.
\ei
The results in Figure \ref{fig:simul_study} suggest that the results are not too sensitive to small-to-moderate misspecifications of the true neighborhoods $\mN_i^\star$.
That said,
if $\mN_i = \mP$ for all units $i \in \mP$ and therefore the misspecification is large,
estimators of edge and spillover terms exhibit bias.

\section{Advanced Topics}
\label{sec:extensions}

We discuss advanced topics,
including model comparisons in Section \ref{sec:model.comparison} and custom-built model terms in Section \ref{sec:iglm.userterms}.

\subsection{Model Comparison}
\label{sec:model.comparison}

While model selection methods with theoretical guarantees for dependent attributes $(X_i, Y_i)$ and connections $Z_{i,j}$ have not been developed,
it is possible to compare models based on predictions.
We assess the added value of the joint probability model for $(\bX, \bY, \bZ)$ specified in Section \ref{sec:specification} over GLMs for $Y_i \mid X_i = x_i$ and $Z_{i,j} \mid (X_i, X_j) = (x_i, x_j)$ based on predictions.

We specify and estimate a simple GLM for $Y_i \mid X_i = x_i$ and $Z_{i,j} \mid (X_i, X_j) = (x_i, x_j)$:
\begin{verbatim}
R> control.obj <- control.iglm(max_it = 300)
R> formula.glm <- data.object ~
+                 attribute_y + 
+                 attribute_xy + 
+                 cov_y(data = white_attribute) +
+                 cov_y(data = gender_attribute) +
+                 edges(mode = "alocal") + 
+                 edges(mode = "local")  +
+                 cov_z(data = match_gender, mode = "local") +
+                 cov_z(data = match_race, mode = "local") +
+                 cov_z(data = match_state, mode = "local") 
R> model.glm <- iglm(formula = formula.glm, 
+                    control = control.obj, 
+                    name = "glm")
R> model.glm$estimate()
\end{verbatim}
The model specified above implies that $Y_i \mid X_i = x_i\, \ind\, \text{Bernoulli}(\mu_{i})$ and $Z_{i,j} \mid (X_i, X_j) = (x_i, x_j)\, \ind\, \text{Bernoulli}(\mu_{i,j})$, 
where 
\beno
\log\dfrac{\mu_{i}}{1-\mu_i}
&\coloneqq& \theta_{\mY} + \theta_{\mX,\mY}\, x_i + \theta_{\mV,\mY,1}\, v_{i,1} + \theta_{\mV,\mY,2}\, v_{i,2}
\ee
and
\beno 
\log\dfrac{\mu_{i,j}}{1-\mu_{i,j}}
&\coloneqq& \theta_{\mZ,1}\, c_{i,j} + \theta_{\mZ,2}\,(1-c_{i,j}) + \dsum_{k=1}^3 \theta_{\mV,\mZ,k}\, \mathbb{I}(v_{i,k} = v_{j,k}).
\ee
We refer to the above model as a GLM,
and to the model specified in Section \ref{sec:specification} as a IGLM.

\s 

\begin{figure}[t!]
    \centering
    \includegraphics[width=0.48\textwidth,clip]{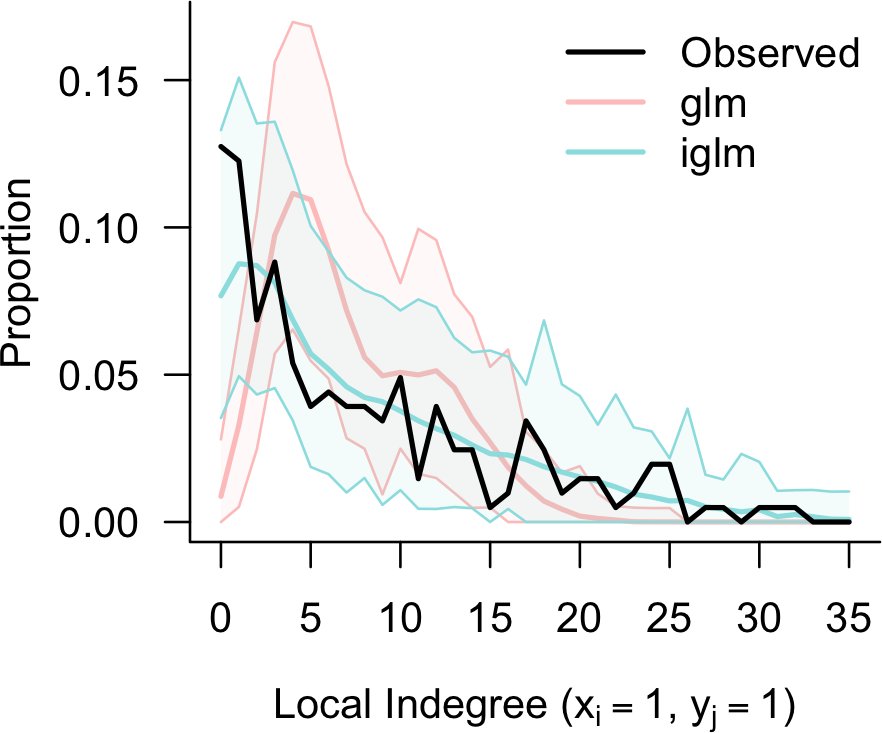}\hfill
    \includegraphics[width=0.48\textwidth,clip]{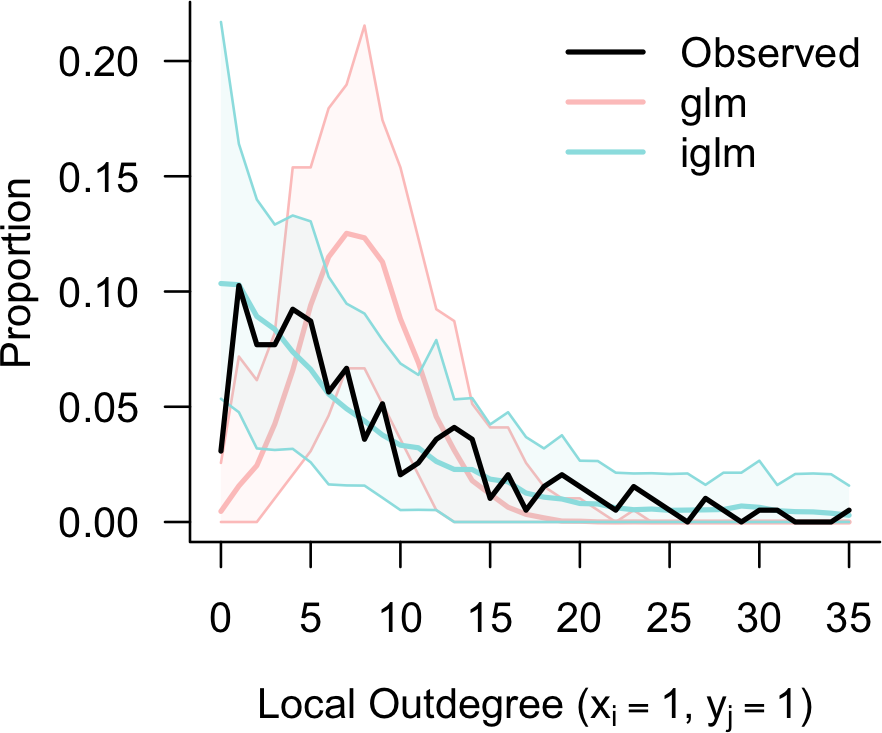}

    \vspace{1em}

    \includegraphics[width=0.48\textwidth,clip]{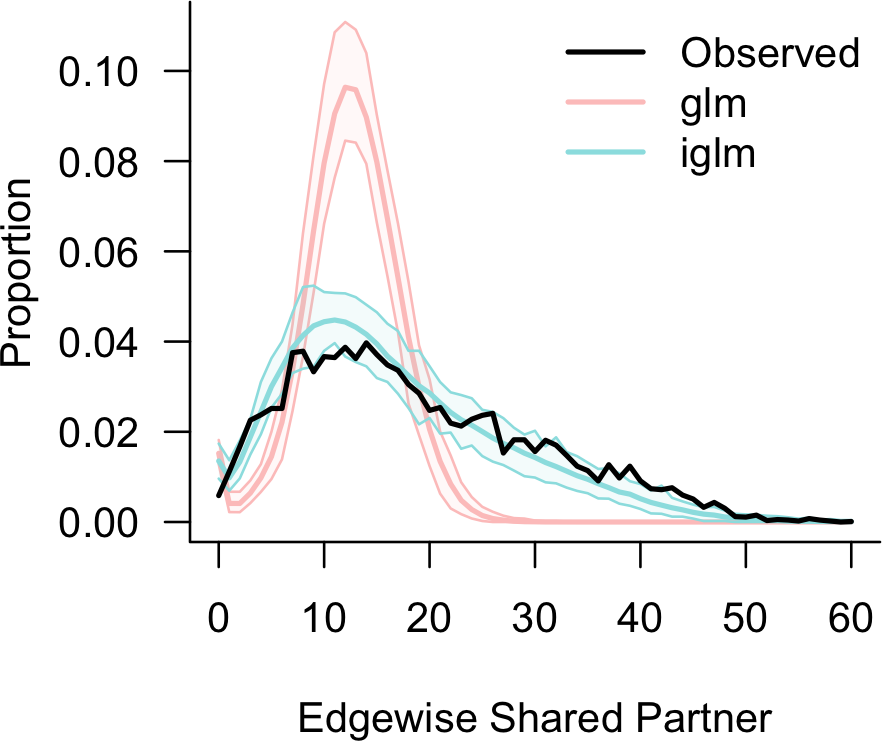}\hfill
    \includegraphics[width=0.48\textwidth,clip]{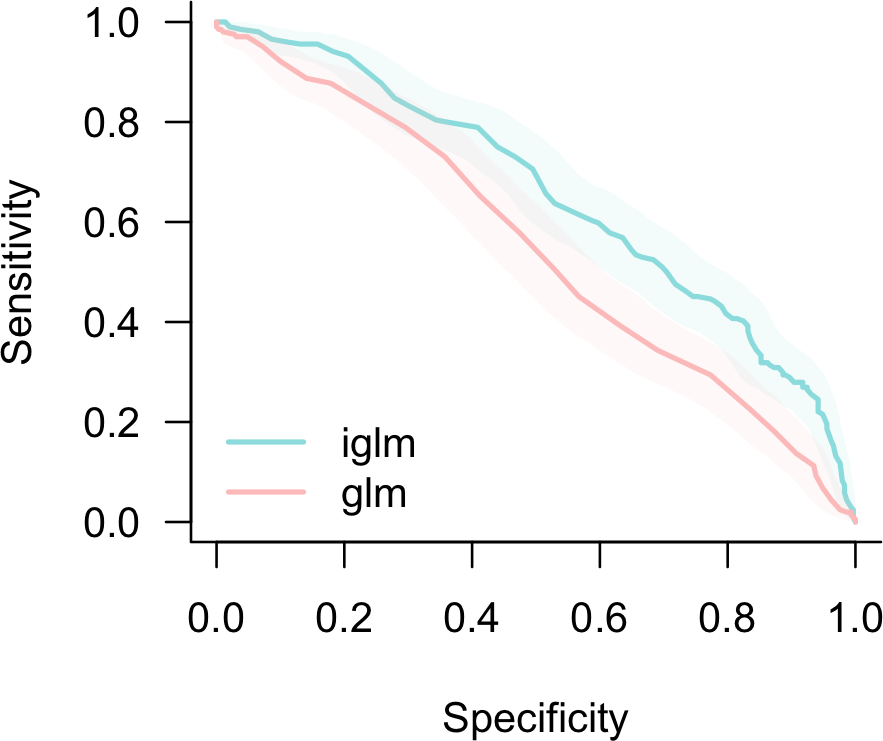}
    \caption{Hate speech on X: 
    Model assessment comparing regression models without interference (GLM) and regression models with interference (IGLM) based on predictions. 
    The bottom-right panel displays Receiver Operating Characteristic (ROC) curves, 
    representing the trade-off between the true positive and false positive rate for outcomes $Y_i$ under both specifications.     
    Shaded regions in the top panels and bottom-left panel indicate the empirical minima and maxima evaluated across all simulations, 
    whereas the shaded region in the bottom-right panel represents a $95\%$-confidence band.
    }
    \label{fig:comparison}
\end{figure}

{\bf Comparison via Assessment Framework}
The diagnostic framework introduced in Section \ref{sec:assessment} enables a visual comparison of models based on predictions of attributes using the function \code{assess()},
suppressing the default graphical output by setting \code{plot = FALSE}:
\begin{verbatim}
R> model.iglm$assess(formula = ~ deg_dist(x_i = 1, y_j = 1, mode = "local") +
+                    esp_dist, plot = FALSE)
R> model.glm$assess(formula = ~ deg_dist(x_i = 1, y_j = 1, mode = "local") +
+                    esp_dist, plot = FALSE)
\end{verbatim}
Using the dots (\code{...}) mechanism of the \code{plot()} method of the \code{results} object,
users can compare predictions:
\begin{verbatim}
R> model.glm$results$plot(model_assessment = TRUE, 
+                         "iglm" = model.iglm$results$model_assessment)
\end{verbatim}
The results are shown in Figure \ref{fig:comparison}.
We interpret these results below,
but we first discuss conditional and marginal predictions of outcomes $Y_i$.
\captionsetup[subfigure]{
    margin={0cm, 1.7cm} 
}

\s 

{\bf Conditional and Marginal Predictions}
The method \code{predict} can calculate conditional and marginal predictions of outcomes $Y_i$ based on approximations of conditional expectations (holding all other variables constant) and marginal expectations (averaging simulations from the model).
To compare observed and predicted quantities,
we leverage established diagnostic frameworks for GLMs:
\begin{enumerate}
\item $Y_i \in \{0, 1\}$:
Receiver Operating Characteristic (ROC) or Precision-Recall curves implemented in \proglang{R} package \pkg{pROC} \citep{pROC}.
\item $Y_i \in \{0, 1, \ldots\}$:
rootograms available in \proglang{R} package \pkg{vcd} \citep{vcd}.
\item $Y_i \in \mR$:
scatterplots comparing predicted and observed outcomes.
\end{enumerate}

{\bf Example}
We compute marginal predictions of attributes $(X_i, Y_i)$ using the \code{predict} method:
\begin{verbatim}
R> prediction.iglm <- model.iglm$predict(variant = "marginal")
R> prediction.glm <- model.glm$predict(variant = "marginal")
\end{verbatim}
The resulting list provides the marginal probabilities of $Y_i = 1$ and $Z_{i,j} = 1$ under IGLM and GLM.
We demonstrate how \proglang{R} package \code{pROC} can be used to compare predictions of outcomes $Y_i$ based on the two models using ROC curves:
\begin{verbatim}
R> library(pROC)
R> roc.iglm <-  roc(prediction.iglm$y$target, prediction.iglm$y$prediction)
R> ciobj.iglm <- ci.se(roc.iglm, specificities = seq(0, 1, l = 100)) 
R> roc.glm <- roc(prediction.glm$y$target, prediction.glm$y$prediction)
R> ciobj.glm <- ci.se(roc.glm, specificities = seq(0, 1, l = 100)) 
R> plot(roc.iglm$specificities, roc.iglm$sensitivities,  col = "#9AE1E3",
+       type = "l", lwd = 2, bty ="l", las = 1, 
+       xlab = "Specificity",ylab = "Sensitivity")
R> plot(ciobj.iglm, type = "shape", col = "#9AE1E366", border = NA, 
+       no.roc = TRUE)
R> lines(roc.glm$specificities, roc.glm$sensitivities,  col = "#FDC7C5", 
+        type = "l", lwd = 2, bty ="l",
+        xlab = "Specificity",ylab = "Sensitivity")
R> plot(ciobj.glm, type = "shape", col = "#FF000022", border = NA, 
+       no.roc = TRUE)
R> legend("bottomleft",legend = c("iglm", "glm"),lty = c(1,1),lwd = c(2,2),
+         col = c("#9AE1E3", "#FDC7C5"), bty = "n")
\end{verbatim}
The results are shown in Figure \ref{fig:comparison}.
These results contain two important lessons:
First,
the fact that model-based predictions of the network based on IGLMs are superior to those based on GLMs suggests that unobserved heterogeneity in the connection propensities of units should be captured along with transitive closure,
using model terms such as \code{degrees} and \code{transitive}.
Second,
the IGLM is superior to GLM in terms of model-based predictions of outcomes,
suggesting that models should include spillover terms to better predict outcomes.

\subsection{User-Defined Model Terms}
\label{sec:iglm.userterms}

Users can add custom-built model terms to the list of model terms implemented and documented in \proglang{R} package \pkg{iglm}.
Since the estimation and simulation algorithms in \pkg{iglm} are implemented in \proglang{C++}, 
custom-built model terms need to be implemented in \proglang{C++}.
To do so, 
users first need to call the function
\begin{verbatim}
R> create_userterms_skeleton()
\end{verbatim}
This function creates an \proglang{R} package called \pkg{iglm.userterms},
which allows users to add custom-built model terms that can be accessed by \pkg{iglm}.   
To add a custom-built model term to \proglang{R} package \pkg{iglm.userterms},
users can follow a simple recipe consisting of four steps:
\begin{enumerate}
\item Decompose $g_i(\cdot)$ and $h_{i,j}(\cdot)$ in accordance with Equation \eqref{eq:lp}.
\item Specify the change statistics $\Delta_{\mX,i}$, 
$\Delta_{\mY,i}$,
and $\Delta_{\mZ,i,j}$ defined in Equations \eqref{eta.eq} and \eqref{change.z}.
\item Write a \proglang{C++} function to compute the change statistics $\Delta_{\mX,i}$, 
$\Delta_{\mY,i}$,
and $\Delta_{\mZ,i,j}$ and register it under the name \code{my_term}.
Store the \proglang{C++} function in a file with extension \code{.cpp} and add it to the \code{src} folder of \pkg{iglm.userterms}.
\item Write a \proglang{R} function called \code{InitIglmTerm.my_term} to check the arguments of the \proglang{C++} function;
note that the extension \code{my_term} of the \proglang{R} function needs to match the name under which the \proglang{C++} function is registered.
Store the \proglang{R} function in the \code{R} folder of \pkg{iglm.userterms}.
\end{enumerate}
To demonstrate,
we implement the treatment spillover term
\be
\label{eq:suff_example}
c_{i,j}\, (x_i^\star\, y^\star_j + x_j^\star\, y_i^\star)\, z_{i,j}
&=& c_{i,j}\, x_i^\star\, y^\star_j\, z_{i,j} 
+ c_{i,j}\, x_j^\star\, y_i^\star\, z_{i,j}
\ee  
assuming that connections are undirected,
so that $z_{i,j} = z_{j,i}$ for all $i < j$.
We walk users through each step of the recipe sketched above:
\begin{enumerate}
\item The treatment spillover term in Equation \eqref{eq:suff_example} is included in the function $h_{i,j}(\cdot)$,
which can be decomposed along the lines of Equation \eqref{eq:lp} as follows:
\beno
h_{\mX,i,j,0}(y_i^\star,y^\star_j,\bz)
&\coloneqq& 0,\;
&
h_{\mX,i,j,1}(y_i^\star,y^\star_j,\bz)
&\coloneqq& \textcolor[HTML]{D55E00}{c_{i,j}\, y^\star_j\, z_{i,j}}\s 
\\
h_{\mX,i,j,2}(y_i^\star,y^\star_j,\bz)
&\coloneqq& c_{i,j}\, y_i^\star\, z_{i,j},\,
&
h_{\mX,i,j,3}(y_i^\star,y^\star_j,\bz)
&\coloneqq& 0\s 
\\
h_{\mY,i,j,0}(x_i^\star,x_j^\star,\bz)
&\coloneqq& 0,\;
&
h_{\mY,i,j,1}(x_i^\star,x_j^\star,\bz)
&\coloneqq& \textcolor[HTML]{0072B2}{c_{i,j}\, x_j^\star\, z_{i,j}}\s
\\
h_{\mY,i,j,2}(x_i^\star,x_j^\star,\bz)
&\coloneqq& c_{i,j}\, x_i^\star\, z_{i,j},
& 
h_{\mY,i,j,3}(x_i^\star,x_j^\star,\bz)
&\coloneqq& 0,
\hide{
\s\s\s
\\
h_{\mX,i,j,0}(x_j^\star, y_i^\star, y^\star_j, \bz) 
&\coloneqq& c_{i,j}\, x_j^\star\, y_i^\star\, z_{i,j},\;
& h_{\mX,i,j,1}(x_j^\star, y_i^\star, y^\star_j, \bz)
&\coloneqq& \textcolor[HTML]{D55E00}{c_{i,j}\, y^\star_j\, z_{i,j}}\s
\\ 
h_{\mY,i,j,0}(x_i^\star, x_j^\star, y^\star_j, \bz) 
&\coloneqq& c_{i,j}\, x_i^\star\, y^\star_j\, z_{i,j},\;
& h_{\mY,i,j,1}(x_i^\star, x_j^\star, y^\star_j, \bz)
&\coloneqq& \textcolor[HTML]{0072B2}{c_{i,j}\, x_j^\star\, z_{i,j}}.
}
\hide{
\s 
\\
h_{\mZ,i,j,0}(\bx, \by, \bz_{-(i,j)}) 
&\coloneqq& 0,\;
& h_{\mZ,i,j,1}(\bx, \by, \bz_{-(i,j)})
&\coloneqq& c_{i,j}\, (x_i\, y_j + x_j\, y_i).
}
\ee
assuming that $i < j$;
note that $h_{\mX,i,j,k}(\cdot) \coloneqq 0$ and $h_{\mY,i,j,k}(\cdot) \coloneqq 0$ ($k = 0, 1, 2, 3$) for all $i > j$,
because we focus on undirected connections.
\item The corresponding change statistics $\Delta_{\mX,i}$, 
$\Delta_{\mY,i}$, 
and $\Delta_{\mZ,i,j}$ are 
\beno 
\Delta_{\mX,i} 
\coloneqq \dsum_{j\in\mP\setminus\{i\}} \textcolor[HTML]{D55E00}{c_{i,j}\, y^\star_j\, z_{i,j}},  
& \Delta_{\mY,i} 
\coloneqq \dsum_{j\in\mP\setminus\{i\}} \textcolor[HTML]{0072B2}{c_{i,j}\, x_j^\star\, z_{i,j}},
& \Delta_{\mZ,i,j}
\coloneqq c_{i,j}\, (x_i^\star\, y^\star_j + y_i^\star\, x_j^\star).
\ee
\item We write the \proglang{C++} function \code{my_stat_spillover} to compute the change statistics $\Delta_{\mathcal{X},i}$, 
$\Delta_{\mY,i}$, 
and $\Delta_{\mZ,i,j}$.
The \proglang{C++} function \code{my_stat_spillover} relies on methods of \proglang{C++} class \code{XYZ_class}, 
which is the \proglang{C++} analog of the class \code{iglm} described in Section \ref{sec:specification}.
Appendix \ref{sec:c++} describes how to use \proglang{C++} class \code{XYZ_class}.
The change statistic $\Delta_{\mX,i}$ is computed by the following \proglang{C++} function on lines 10--15, 
$\Delta_{\mY,i}$ is computed on lines 16--21, 
and $\Delta_{\mZ,i,j}$ is computed on lines 22--29:
\begin{Verbatim}[numbers=left, stepnumber=1, numbersep=5pt, firstnumber=1]
double my_stat_spillover(const XYZ_class &object,
                         const int &unit_i,
                         const int &unit_j,
                         const arma::mat &data,
                         const double &type,
                         const std::string &mode,
                         const bool &is_full_neighborhood)
{
  double res = 0.0;
  if (mode == "x") { // x_i from 0 -> 1
    const auto& connections_of_i = object.adj_list_nb.at(unit_i);
    for (const int &k : connections_of_i) {
      res += object.y_attribute.get_val(k);
    }
  } 
  else if (mode == "y") { // y_i from 0 -> 1
    const auto& connections_of_i = object.adj_list_nb.at(unit_i);
    for (const int &k : connections_of_i) {
      res += object.x_attribute.get_val(k);  
    }
  } 
  else { // z_ij from 0 -> 1
    if (object.get_val_overlap(unit_i, unit_j)) {
      res = (object.x_attribute.get_val(unit_i) * 
             object.y_attribute.get_val(unit_j)) +
            (object.x_attribute.get_val(unit_j) * 
             object.y_attribute.get_val(unit_i));
    }
  }
  return res;
}
EFFECT_REGISTER("my_spillover", ::my_stat_spillover, "my_spillover", 0);
\end{Verbatim}
The \proglang{C++} function \code{EFFECT_REGISTER} registers the created term under the name\break 
\code{my_spillover} on line 32.
The last argument of \code{EFFECT_REGISTER} specifies the value of the statistic when no connections are observed,
which is $0$.
We store the \proglang{C++} function \code{my_stat_spillover} in the \code{src} folder of \proglang{R} package \pkg{iglm.userterms}.
\item Since the \proglang{C++} function was registered under the name \code{my_spillover}, we define the corresponding \proglang{R} function \code{InitIglmTerm.my_spillover} as follows: 
\begin{verbatim}
R> InitIglmTerm.my_spillover <- function(data_object, arglist, ...) {
+    arglist <- iglm:::check.IglmTerm(data_object, arglist,
+                                     directed = FALSE)
+    list(
+         term_name = "my_spillover",
+         coef_name = arglist$label
+    )
+  }
\end{verbatim}
We store the \proglang{R} function \code{InitIglmTerm.my_spillover} in the \proglang{R} folder of \pkg{iglm.userterms}.
\end{enumerate}
The final step is to compile \proglang{R} package \pkg{iglm.userterms} and load it after \pkg{iglm}:
\begin{verbatim}
R> library(iglm)
R> library(iglm.userterms)
\end{verbatim}
We can then use the term \code{my_spillover}:
e.g.,
\code{statistics(data.object ~ my_spillover)} counts $\sum_{(i,j) \in\mathscr{D}} c_{i,j}\, (x_i^\star\, y^\star_j + x_j^\star\, y_i^\star)\, z_{i,j}$.
Users interested in adding custom-built model terms can inspect the \proglang{C++} file \code{change_statistics.cpp} in the \proglang{C++} source folder of \proglang{R} package \pkg{iglm},
which contains \proglang{C++} implementations of all model terms listed in the manual of \pkg{iglm}.

\section{Application: Real-Valued Outcomes}
\label{sec:application}

In addition to the running example with binary outcomes $Y_i \in \{0, 1\}$,
we provide an example with real-valued outcomes $Y_i \in \mR$.
To do so,
we use the Copenhagen network, 
which investigates communications among $N = 409$ students at a university in Copenhagen, 
Denmark over 28 days.
The data collection procedure is described in \citet{sapiezynskiInteractionDataCopenhagen2019} and involved recording interaction metrics, 
including Bluetooth-based proximity scans and self-reported friendships, 
where $Z_{i,j} \coloneqq 1$ indicates that $i$ considers $j$ to be a friend or $j$ considers $i$ to be a friend,
and $Z_{i,j} \coloneqq 0$ otherwise.
In other words,
the network is undirected.
The fixed predictor $X_i \in \{0, 1\}$ encodes gender ($1$ if female and $0$ otherwise).
The outcome $Y_i \in \mR$ represents the log-transformed total call duration $Y_i \coloneqq \log(t_i)$, 
where $t_i$ is the total number of minutes spent on phone calls;
note that $t_i = 0$ is impossible because the availability of call information was one of the inclusion criteria in the study.
The neighborhood of student $i$ is defined as
\beno
\mN_i 
&\coloneqq& \{j \in \mP: \text{\em{ $j$ was detected by $i$'s Bluetooth device for at least $24$ hours}}\}.
\ee
The preprocessed data object is included in \proglang{R} package \code{iglm} as an example data set,
and we can load and plot it as follows:
\begin{verbatim}
R> data("copenhagen")
R> set.seed(321)
R> copenhagen$plot()
\end{verbatim}
\begin{figure}[t!]
  \centering
  \includegraphics[width=0.7\textwidth, clip]{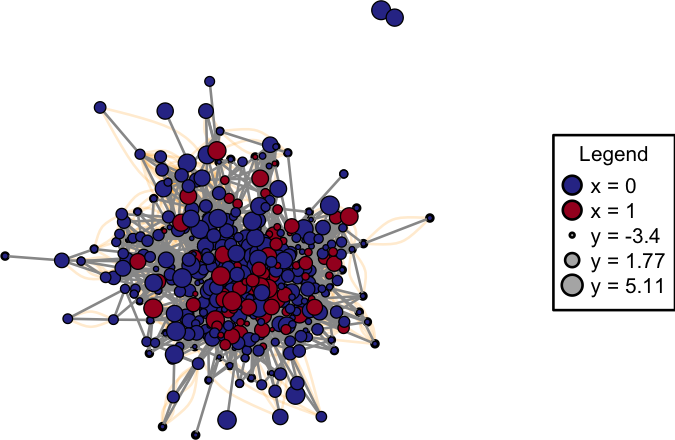}
   \caption{  
   \label{fig:igraph}
   Copenhagen network:
   The colors and sizes of the circles represent the gender $X_i$ and log-transformed call time $Y_i$ of students $i$, 
   respectively.
   }
  \label{fig:plot_copenhagen}
\end{figure}
The function \code{plot()} plots the network in Figure \ref{fig:plot_copenhagen}.
Since $Y_i \in \mR$,
\code{attribute_y} is set to \code{"normal"}, 
and the conditional variance $\psi_\mathscr{Y}$ of $Y_i$  is estimated by its empirical variance:
\begin{verbatim}
R> copenhagen$set_scale_y(var(copenhagen$y_attribute))
\end{verbatim}
Users can specify labels \code{label_x, label_y, label_z}\, of attributes and connections \code{x_attribute, y_attribute, z_network}.
The labels are specified as follows:
\begin{verbatim}
R> copenhagen$set_label_x("gender")
R> copenhagen$set_label_y("duration")
R> copenhagen$set_label_z("friendship")
\end{verbatim}
These labels are used in descriptive statistics and other output:
\begin{verbatim}
R> copenhagen
iglm.data object
  units                                   : 409
  directed                                : FALSE
  connections [friendship] (fixed = FALSE): 2517
  local dyads                             : 675 / 83436 (0.8%)
Attribute summaries
  x_attribute [gender] (fixed = TRUE): binomial mean = 0.22
  y_attribute [duration]      : normal mean = 1.33, sd = 1.78
\end{verbatim}
We then specify the model and estimate it:
\begin{verbatim}
R> control.obj <- control.iglm(max_it = 300) 
R> formula.norm <- copenhagen ~
+                  attribute_y +
+                  attribute_xy +
+                  degrees + 
+                  edges(mode = "alocal") +
+                  transitive + 
+                  spillover_xy +
+                  spillover_yy 
R> model.norm <- iglm(formula = formula.norm, control = control.obj)
R> model.norm$estimate()
\end{verbatim}
\hide{
The specified model includes spillover terms,
which are local by default.
Since none of the students has more than 10 neighbors,
we use the unscaled spillover terms.
}
The results are:  
\begin{verbatim}
R> model.norm$summary()
iglm object
------------------------------------------------------------
Results: 

                            Estimate   S.E. t-value Pr(>|t|)
attribute_duration             1.082  0.117     9.2 < 0.0001
attribute_gender_duration      0.420  0.259     1.6  0.10416
edges(mode = 'alocal')        -1.246  0.230    -5.4 < 0.0001
transitive                     2.810  0.245    11.5 < 0.0001
spillover_gender_duration     -0.245  0.136    -1.8  0.07241
spillover_duration_duration    0.342  0.096     3.6  0.00038

Time for estimation: 14 secs

Degree Parameters:
   Min. 1st Qu.  Median    Mean 3rd Qu.    Max. 
  -4.51   -1.99   -1.51   -1.51   -1.00    0.77 
\end{verbatim}
The function \code{summary()} uses the labels of attributes and connections specified above.

The conditional distribution of outcome $Y_i \in \mR$ is Normal with mean
\beno
\mu_{\mY,i}(\widehat\eta_{\mY,i})
&\coloneqq& 1.082 + .420\, x_{i}^\star - .245 \dsum_{j \in \mP \setminus \{i\}}\, c_{i,j}\, x_j^\star\, z_{i,j} + .342 \dsum_{j \in \mP \setminus \{i\}}\, c_{i,j}\, y^\star_j\, z_{i,j}.
\ee
Note that the linear predictor differs from the linear predictor in the running example in Section \ref{sec:model},
because $z_{i,j} = z_{j,i}$ and $h_{j,i}(\cdot) \coloneqq 0$ ($i < j$).
While the negative treatment spillover effect is not significant, 
the positive outcome spillover effect is significant at conventional significance levels (e.g., $.05$).
In other words,
the more time $i$'s friend $j$ spends on phone calls,
the more time $i$ is expected to spend on phone calls.

The conditional distribution of connection $Z_{i,j} \in \{0, 1\}$ is Bernoulli with mean
\beno
\mu_{\mZ,i,j}(\widehat\eta_{\mZ,i,j})
\;\coloneqq\; \mbE_{\widehat\eta_{\mZ,i,j}}[Z_{i,j} \mid (\bX, \bY, \bZ_{-(i,j)}) = (\bx, \by, \bz_{-(i,j)})]
\;=\; \dfrac{\exp(\widehat\eta_{\mZ,i,j})}{1 + \exp(\widehat\eta_{\mZ,i,j})},
\ee
where
\beno 
\widehat\eta_{\mZ,i,j}
\coloneqq \widehat\theta_{\mZ,i} + \widehat\theta_{\mZ,j} -1.246\, (1 - c_{i,j}) + c_{i,j}\, \left[2.810 \, \Delta_{\mZ,i,j}(\bz)
- .245\, (x_i^\star\, y^\star_j + x_j^\star\, y_i^\star)
+ .342\, y_i^\star\, y^\star_j \right],
\ee
and $\widehat\theta_{\mZ,i}$ and $\widehat\theta_{\mZ,j}$ can be interpreted as the propensities of students $i$ and $j$ to connect with others.
Here,
too,
there is strong evidence for transitive closure among friends:
More often than not,
a friend of a friend is a friend.

\begin{figure}[t!]
  \centering
  \includegraphics[width=0.45\textwidth, clip]{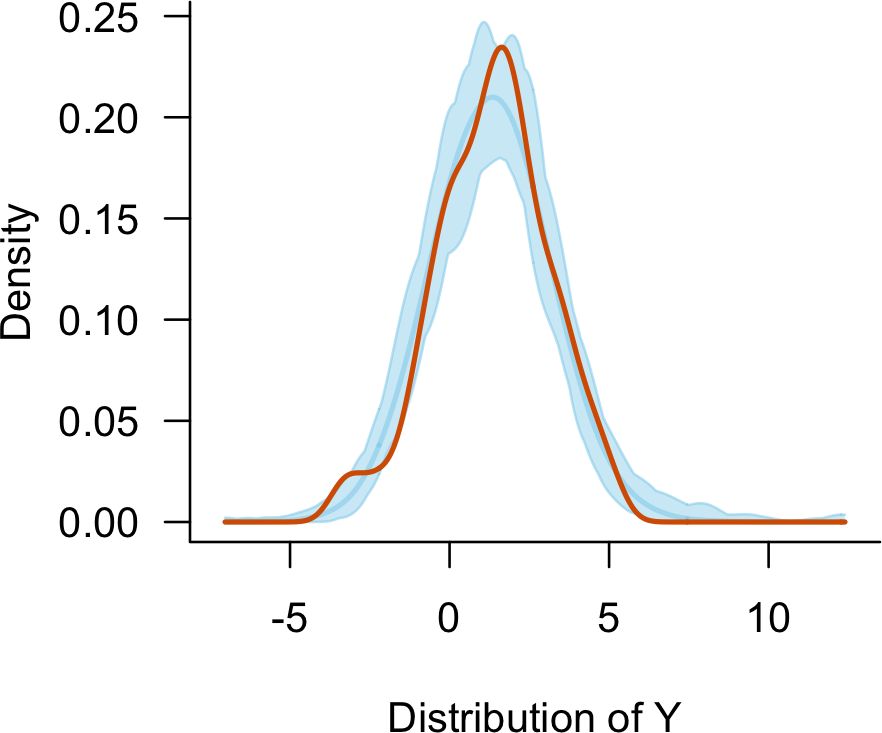}
  \includegraphics[width=0.45\textwidth, clip]{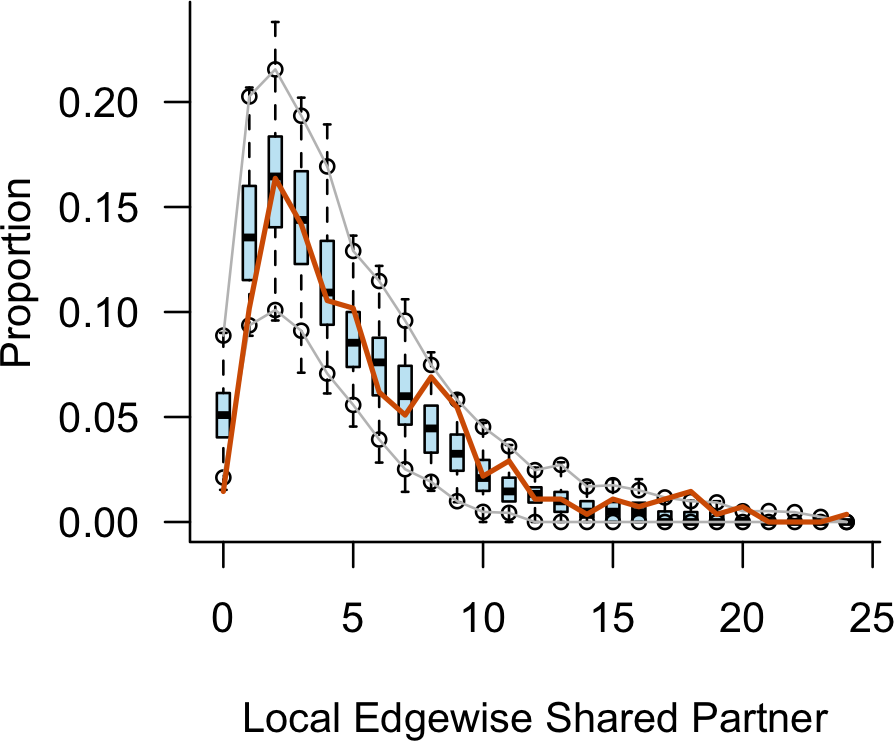} 
  
  \s\s 
  
  \includegraphics[width=0.45\textwidth, clip]{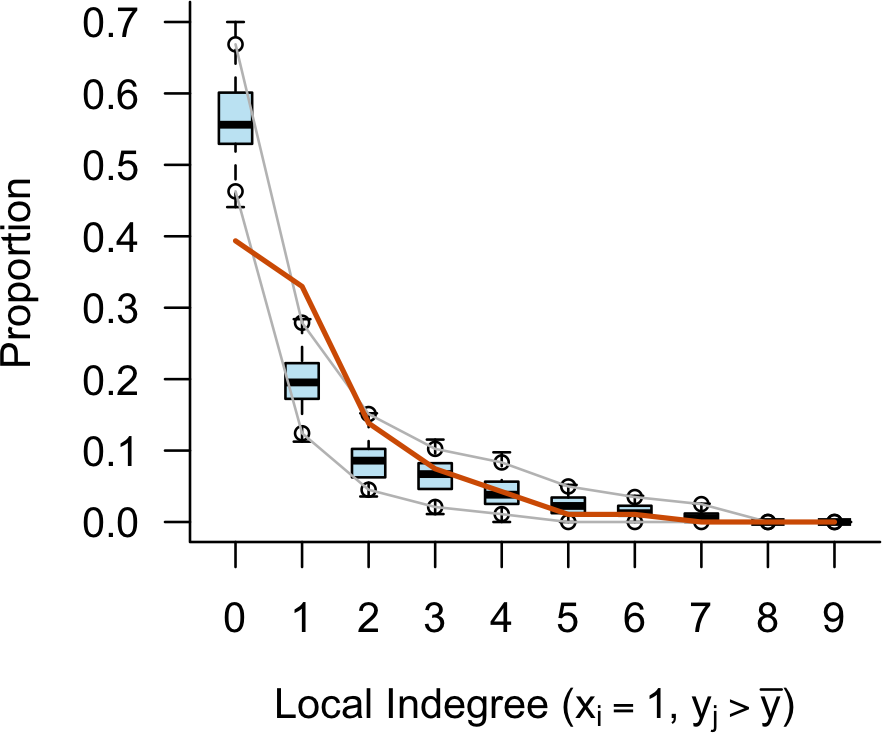}
  \includegraphics[width=0.45\textwidth, clip]{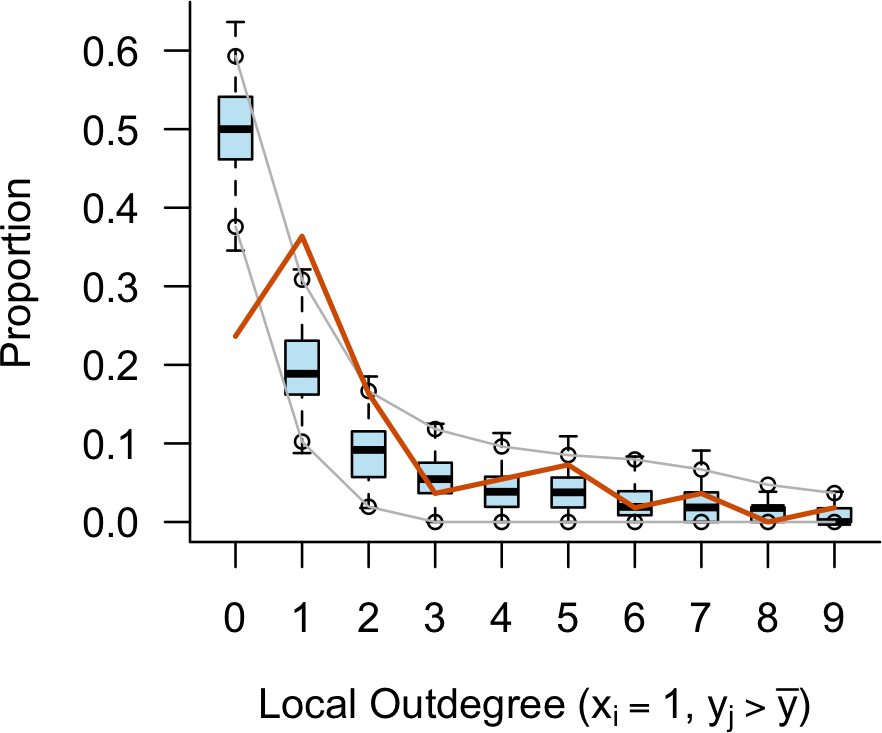}
   \caption{Copenhagen network:
   Model assessment in terms of outcomes $Y_i$,
   the number of local edgewise shared partners among pairs of units $\{i, j\}$ with $c_{i,j} = 1$,
   and local spillover in- and outdegrees among pairs of units $\{i, j\}$ with $c_{i,j} = 1$, $x_i = 1$, and $y_j > \bar{y}$.
   }
  \label{fig:model_assessment_copenhagen}
\end{figure}

Finally,
we assess the model by comparing simulated and observed outcomes $Y_i$ as well as spillover in- and outdegrees and geodesic distance distributions: 
\begin{verbatim}
R> model.norm$assess(formula = ~  y_dist + esp_dist(mode = "local") +
+                                 deg_dist(x_i = 1, y_j = 1, mode = "local"))
\end{verbatim}
Figure \ref{fig:model_assessment_copenhagen} suggests that the model matches the observed data in terms of the marginal distribution of outcomes $Y_i$,
local spillover in- and outdegrees,
and local geodesic distances.

\section{Discussion}
\label{sec:discussion}

We conclude with a discussion of open questions.

First,
we have assumed that the neighborhoods of units are either known or can be covered by supersets,
and we have demonstrated that maximum pseudo-likelihood estimators are robust against small-to-moderate misspecifications of neighborhoods.
That said,
sometimes it is impossible or undesirable to specify neighborhoods,
in which case one may wish to learn neighborhoods from data.
Neighborhoods can be learned from observations of
\bi
\item[(a)] attributes $(\bX, \bY)$;
\item[(b)] connections $\bZ$;
\item[(c)] both attributes $(\bX, \bY)$ and connections $\bZ$.
\ei 
Approach (a) can leverage methods developed for high-dimensional graphical models \citep[e.g.,][]{MeBu06,RaWaLa10},
but existing justifications for these methods rely on independent replications that are unavailable in connected populations with dependent attributes $(\bX, \bY)$ and connections $\bZ$.
Approach (b) can take advantage of stochastic block models and other clustering methods \citep[e.g.,][]{Ab18,GaMa21},
but most of those approaches assume that connections are independent conditional on communities.
We are not aware of methods implementing approach (c),
which is the most promising approach.
How to learn unknown neighborhoods from data without independent replications is an important direction for future research.

Second,
\proglang{R} package \pkg{iglm} reports tail probabilities \code{Pr(>|t|)},
in line with other \proglang{R} packages.
In general,
tail probabilities can be obtained from
\bi
\item normal distributions,
supported by Berry-Esseen-type bounds for normal approximations \citep[e.g.,][]{St24} or Central Limit Theorems \citep[e.g.,][]{ScHa13},
both of which focus on dependent connections $Z_{i,j}$ with local dependence within non-overlapping subpopulations;
\item non-asymptotic concentration inequalities for dependent random variables \citep[e.g.,][]{StSc20},
which allow the dependence of connections $Z_{i,j}$ to propagate throughout the population but require controlling dependence using,
e.g., 
neighborhoods that overlap but are not too large and do not overlap too much.
\ei 
The cited works provide tail probabilities based on connections $\bZ$ while ignoring attributes $(\bX, \bY)$.
Extensions to count- and real-valued attributes $(\bX, \bY)$ are non-trivial,
but simulation results in \citet{FrScBhHu24} suggest that the finite-sample distributions of maximum pseudo-likelihood estimators are approximately normal.
\proglang{R} package \pkg{iglm} therefore reports tail probabilities \code{Pr(>|t|)} based on $N(0, 1)$ distributions with mean $0$ and variance $1$.
Berry-Esseen-type bounds for normal approximations of estimators based on count- and real-valued attributes $(\bX, \bY)$ and connections $\bZ$ is an important direction for future research.

\section*{Acknowledgments}

We acknowledge support by U.S.\ National Science Foundation award NSF DMS-2515763 and U.S.\ Army Research Office award ARO W911NF-21-1-0335.
We are indebted to two referees whose constructive suggestions have greatly improved the manuscript.
We are grateful to Subhankar Bhadra and Donghwi Nam for valuable feedback on drafts of this manuscript.

\bibliography{base}

\newpage

\begin{appendix}

\section{Details on the C++ Implementation}
\label{sec:c++}

The \proglang{C++} class \code{XYZ_class} is analogous to the \code{R6} class \code{iglm.data} in \proglang{R} with the following main attributes: 
\begin{itemize}
    \item \code{int n_units}: 
    number of units; 
    \item \code{Attribute x_attribute} and \code{Attribute y_attribute}: 
     representations of $\bx$ and $\by$, where \code{Attribute} is a lightweight class acting as a wrapper of the \code{n_units}-dimensional vector internally represented as an \code{arma::vec} object; 
    \item \code{std::vector<std::vector<int>\,> z_network.adj_list} and \newline  \code{std::vector<std::vector<int>\,> z_network.adj_list_in}: representation of $\bz$ via two edge lists: one for outgoing and one for ingoing connections;
    in case of undirected connections, \code{z_network.adj_list_in} is not populated and all connections are symmetrized in \code{z_network.adj_list}; 
    \item \code{std::vector<std::vector<int>\,> neighborhood}: 
    representation of $\mN_i$ for $i = 1, ..., N$ via a neighborhood list; \item \code{std::vector<std::vector<int>\,> overlap}: 
    representation of $\mathscr{O}_i \coloneqq \{j: \mN_i \cap \mN_j \neq \emptyset\}$ for $i = 1, ..., N$ via an overlap list; 
    \item \code{std::vector<std::vector<int>\,> adj_list_nb} and \newline \code{std::vector<std::vector<int>\,> adj_list_in_nb}: 
    two edge lists for connections with overlapping neighborhoods: one for outgoing and one for ingoing connections. In case of undirected connections, \code{adj_list_in_nb} is not populated and all connections are symmetrized in \code{adj_list_nb}; 
\end{itemize}
The only occasion for users to interact with the \proglang{C++} backend is when defining a function to compute the change statistics $\Delta_{\mX,i}$, 
$\Delta_{\mY,i}$,
and $\Delta_{\mZ,i,j}$ for additional terms.
In this setting,
the \code{XYZ_class} instance internally generated in package \pkg{iglm} to match exactly the current state of the provided \code{iglm.data} object 
is always named \code{object}, 
as shown in Section \ref{sec:extensions}. 
Users can, 
for instance, 
access the outgoing edge list for overlapping neighborhoods via \code{object.adj_list_nb}. 
Below is an exhaustive enumeration of the querying methods for this instance users can rely on to compute $\Delta_{\mX,i}, \Delta_{\mY,i},$ and $\Delta_{\mZ,i,j}$. 
First, we list commands to evaluate specific parts of the data:
\begin{itemize}
        \item \textbf{Attribute without scale:}
        To get the value $x_{i}$ or $y_{i}$, call \newline \code{object.x_attribute.get_val_no_scale(i)} or \newline \code{object.y_attribute.get_val_no_scale(i)}, respectively, for \code{int i}.
        To be consistent with \proglang{R}, we implemented 1-based indexing throughout the \proglang{C++} backend. 
        Therefore, \newline \code{object.y_attribute.get_val_no_scale(1)} returns $y_1$, which corresponds to the \code{0}th entry of the underlying \code{arma::vec} object. 
    \item \textbf{Attribute with scale:} 
    To get the value of $x_i^\star \coloneqq x_i/\psi_\mathscr{X}$ or $y_i^\star \coloneqq y_i/\psi_\mathscr{Y}$, call \newline \code{object.x_attribute.get_val(i)} or \code{object.y_attribute.get_val(i)}, respectively, for \code{int i}.
    \item \textbf{Connections:} 
    To get the value of $z_{i,j}$,  call \code{object.z_network.get_val(i, j)} for \code{int i} and \code{int j}.
    \item \textbf{Neighborhood Membership:} 
    To check whether $j \in \mN_i$, call \newline \code{object.get_val_neighborhood(i, j)} for \code{int i} and \code{int j}.
    \item \textbf{Neighborhood Overlap:} 
     To get the value of $c_{i,j} \coloneqq \mathbb{I}(\mN_i \cap \mN_j \neq \emptyset)$, call \newline \code{object.get_val_overlap(i, j)} for \code{int i} and \code{int j}.
\end{itemize}
Second, 
we list additional functions to ease defining custom \proglang{C++} functions: 
\begin{itemize}
    \item \textbf{Out- and Ingoing Connections:} 
    The \code{std::vector<int>} object of all connections from or to unit $i$ can be accessed via \code{object.z_network.adj_list[i]} and \newline \code{object.z_network.adj_list_in[i]}, respectively, for \code{int i}.
    \item \textbf{Connections with Overlapping Neighborhoods:} 
    Connections of unit $i$ restricted to units with overlapping neighborhoods can be accessed via \code{object.adj_list_nb[i]} and \code{object.adj_list_in_nb[i]}, respectively.
    \item \textbf{Units with Overlapping Neighborhoods:} 
    The set of units that have an overlapping neighborhood with unit $i$ can be accessed via \code{object.overlap[i]}.
    \item \textbf{Out- and Indegree:}
    The out- and indegree of unit $i$ are given by \newline \code{object.z_network.out_degrees.at(i)} and\newline  \code{object.z_network.in_degrees.at(i)}, respectively.
    \item \textbf{Out- and Indegree with Overlapping Neighborhood:} 
    The out- and indegree of unit $i$ with other units that have some overlapping neighborhood are given by \newline \code{object.out_degrees_nb.at(i)} and\newline  
    \code{object.in_degrees_nb.at(i)}, 
    respectively.
    \item \textbf{Shared Partners:} 
    To get the set of common partners $k$ connecting $i$ and $j$ in the global network $\bz$ as a \code{std::vector<int>} object, 
    call\newline  \code{object.get_common_partners(i, j, type)} for \code{int i}, \code{int j}, and \code{std::string type}.
     The \code{type} argument specifies the geometric path (e.g., ``OTP'' for Outgoing-Two-Paths where $Z_{ik}\,Z_{kj}=1$, see the \pkg{iglm} manual for full definitions of the other options: ``OSP'' for Outgoing-Shared-Partner, ``ISP'' for Ingoing-Shared-Partner, ``ITP'' for Ingoing-Two-Paths). 
     The same \code{type} argument is used for all ensuing methods. 
    \item \textbf{Shared Partners with Overlapping Neighborhood:} 
    To get the set of common partners $k$ with $c_{i,k} \, c_{j,k}=1$ connecting $i$ and $j$ as a \newline 
    \code{std::vector<int>} object,  call  \code{object.get_common_partners_nb(i, j, type)}.
    \item \textbf{Count Shared Partners:} 
    To compute the number of common partners $k$ connecting $i$ and $j$ in the global network $\bz$, 
    call \newline  \code{object.count_common_partners(i, j, type)}. 
    \item \textbf{Count Shared Partners with Overlapping Neighborhood:} 
    To compute the number of common partners $k$ with $c_{i,k} \, c_{j,k}=1$ connecting $i$ and $j$,  call \newline 
    \code{object.count_common_partners_nb(i, j, type)}.
\end{itemize}

\newpage 

\section{Simulations: Misspecification of Neighborhoods}
\label{sec:simulation_study}

We assess the robustness of maximum pseudo-likelihood estimator $\widehat\nat$ to misspecifications of the neighborhoods using simulations.
We consider a population $\mP \coloneqq \{1, \ldots, 100\}$ consisting of 10 non-overlapping subpopulations $\{1, \ldots, 10\}, \ldots, \{91, \ldots, 100\}$ of size 10.
For each unit $i \in \mP$,
the true neighborhood $\mN_i^\star$ consists of all units in the same subpopulation,
including $\{i\}$.
Throughout,
the specified neighborhood of unit $i \in \mP$ is $\mN_i$,
while the true neighborhood of unit $i \in \mP$ is $\mN_i^\star$.
We consider the following forms of misspecification:
\bi
\item Full neighborhood $\mN_i = \mP$.
\item Supersets $\mN_i \supset \mN_i^\star$,
i.e.,
the specified neighborhoods $\mN_i$ consist of all true neighbors $j \in \mN_i^\star$ and 5\%,
10\%,
and 15\% non-neighbors $j \in \mP \setminus \mN_i^\star$.
\item Subsets $\mN_i \subset \mN_i^\star$,
i.e.,
the specified neighborhoods $\mN_i$ are obtained by taking the set of all true neighbors $j \in \mN_i^\star$ and removing 5\%,
10\%,
and 15\% of the true neighbors $j \in \mN_i^\star$.
\item Misclustering,
i.e.,
we sample at random $5\%, 10\%,$ or $15\%$ of units and move each of them from their home subpopulation to some other subpopulation.
\ei
\hide{
\bi 
\item[]1. Correct specification: $\mN_i = \mN_i^\star$.
\item[]2--4. Supersets $\mN_i \supset \mN_i^\star$: Sample at random $5\%, 10\%,$ or $15\%$ of false-positive neighbors in addition to the true neighbors.
\item[]5--7. Subsets $\mN_i \subset \mN_i^\star$: Sample at random $5\%, 10\%,$ or $15\%$ of true neighbors and remove them from the neighborhood.
\item[]8--10. Misclassification $\mN_i \neq \mN_i^\star$: Sample at random $5\%, 10\%,$ or $15\%$ of units and move each of them from their home subpopulation to some other subpopulation.
\item[]11. Full neighborhood with $\mN_i = \mP$.
\ei 
}
We focus on binary attributes $(\bX, \bY) \in \{0, 1\}^{2\, N}$ and specify
\[
\begin{array}{ccccccc}
g_i(x_i, y_i) 
&=& 
\left(
\begin{array}{c}
x_i\s 
\\
y_i\s 
\\
x_i\,y_i
\end{array}
\right),
&&
\nat_g^\star
&=& 
\left(
\begin{array}{c}
0.5\s 
\\
-1.0\s 
\\
1.2
\end{array}
\right)\s\s
\\
h_{i,j}(x_i, x_j, y_i, y_j, \bz)
&=& 
\left(
\begin{array}{c}
c_{i,j}\, z_{i,j}\s 
\\
(1-c_{i,j})\, z_{i,j}\s
\\
c_{i,j}\,y_i\, y_j\,z_{i,j}
\end{array}
\right),
&&
\nat_h^\star
&=& 
\left(
\begin{array}{c}
0.6\s 
\\
-1.6\s 
\\
0.3
\end{array}
\right).
\end{array}
\]
Under the complete neighborhood specification $\mN_i = \mP$,
the weight of $(1 - c_{i,j})\, z_{i,j}$ is not identifiable,
because
\beno 
(1-c_{i,j})\, z_{i,j} &=& 0 
&\mbox{for all}& 
(i, j) \in \mP^2.
\ee 
We therefore set the weight of $(1 - c_{i,j})\, z_{i,j}$ to $0$ when $\mN_i = \mP$.

In each scenario,
we generate $R = 50$ replications.
The source code to replicate the simulation study is included in the replication script.

\end{appendix}
\end{document}